\documentclass{PoS}

\usepackage{psfrag}

\title{Progress in Kaon Physics on the Lattice}

\ShortTitle{Progress in Kaon Physics on the Lattice}

\author{\speaker{Weonjong Lee} \\
  Frontier Physics Research Division and
  Center for Theoretical Physics, \\
  Department of Physics and Astronomy, \\
  Seoul National University, \\
  Seoul, 151-747, South Korea \\
  E-mail: \email{wlee@phya.snu.ac.kr}}

\abstract{
We review recent progress in calculating kaon spectrum,
pseudoscalar meson decay constants, $B_K$, $\epsilon'/\epsilon$,
$K\rightarrow \pi\pi$ matrix elements, kaon semileptonic form factors,
and moments of kaon distribution amplitudes on the lattice.
We also address the issue of how best to improve the staggered fermion
formulation for the action and operators.
}

\FullConference{XXIV International Symposium on Lattice Field Theory\\
		 July 23-28 2006\\
		 Tucson Arizona, US}

\begin{document}

\section{Introduction}
\label{sec:intro}
Kaon physics is a rich resource when searching for new physics beyond
the standard model.
There have been a series of essential quantum jumps in the
understanding of kaon spectrum, kaon decays, kaon mixing, partonic
analysis of kaon, kaon form factors, and so on.
In this article, we will review some of the recent progress in kaon
physics.

First, we review recent progress in understanding the pattern of taste
symmetry breaking in the pion multiplet spectrum of staggered fermions.
We compare a number of popular improvement schemes for staggered
fermions and provide a guide to the best scheme by comparing the
mass splittings of the pion multiplet in various improvement schemes.

Second, we address the issue of calculating $f_\pi$ and $f_K$
with extremely high precision, which have a massive impact in
determining some of the low energy constants such as $L_5$ and the CKM
matrix elements.
We review the results of MILC (AsqTad staggered fermions in $2+1$
flavor QCD), RBC/UKQCD (domain wall fermions in $2+1$ flavor QCD),
and NPLQCD (a mixed action scheme in $2+1$ flavor QCD).

Third, we review the phenomenological importance of indirect CP
violation for the CKM unitarity triangle.
We review the history of calculating $B_K$ in quenched QCD,
and compare the capabilities of the various improvement schemes
for staggered fermions to reduce scaling violations in $B_K$.
This comparison will guide us to the best scheme to improve staggered
fermions.
We will also review recent progress in calculating $B_K$ in $2+1$
flavor QCD using improved staggered fermions and domain wall fermions.

Fourth, we address critical issues in calculating $\epsilon'/\epsilon$
on the lattice.
We explain a serious ambiguity in defining a lattice version of
left-right QCD penguin operators, in particular $Q_6$, which was
originally pointed out by Golterman and Pallante.
We review a recent numerical study on this issue by RBC, which confirms
the original prediction by Golterman and Peris.
This leads to a conclusive claim that it is not possible to reliably
calculate the leading contribution of $Q_6$ to $\epsilon'/\epsilon$ in
quenched QCD.
In addition, we also review theoretical progress in understanding the
left-left QCD penguin operators by Golterman and Pallante.
This explains the existence of another serious ambiguity in a lattice
version of left-left QCD penguin operators, in particular $Q_1$, $Q_2$
and $Q_4$, which leads to the conclusion that it is not possible to
reliably calculate even the sub-leading contribution of $Q_4$ to
$\epsilon'/\epsilon$ in quenched QCD.
We also propose a solution to the problems in the end.

Fifth, we review the history of calculating the hadronic matrix
elements of $K\rightarrow \pi\pi$ directly.
We discuss the Maiani Testa no-go theorem, and consider a number of
methods to get around it: the diagonalization method, the H parity
boundary condition method, and the moving frame method.
We also present recent numerical work in the moving frame by RBC.

Sixth, we review the progress in calculating the vector form factor of
$K_{l3}$ decays on the lattice, which determines $|V_{us}|$ with high
accuracy.
The results of RBC, HPQCD/FNAL, and JLQCD are presented.

Seventh, we review the progress in calculating the first moment of the
kaon distribution amplitude.
The results of UKQCD and QCDSF/UKQCD are presented.

Finally, we summarize the whole discussion and end with some
conclusions.
In the final remarks, we propose to use $\overline{\rm Fat7}$
staggered fermions as the best candidate to generate $2+1$ unquenched
gauge configurations using the rational hybrid Monte Carlo algorithm
(RHMC).

\section{Pion, Kaon spectrum}
\label{sec:spec}
Staggered fermions have four degenerate tastes by construction, which
leads to SU(4) taste symmetry in the continuum limit $a \to 0$.
However, the taste symmetry is broken at a finite lattice spacing
($a>0$).
The splittings in the pion multiplet spectrum are a non-perturbative
measure of the taste symmetry breaking.
From a theoretical point of view, a natural tool to study the pion 
multiplet spectrum is staggered chiral perturbation theory
\cite{ref:wlee:1,ref:bernard:1}.
According to staggered chiral perturbation theory, the taste
symmetry breaking in pion multiplet spectrum happens in two steps: at
${\cal O}(a^2) \approx {\cal O}(p^2)$ the SU(4) taste symmetry is
broken down to SO(4) taste symmetry, and at higher order of ${\cal
O}(a^2 p^2)$ the SO(4) taste symmetry is broken down to the discrete
spin-taste symmetry $SW_4$ \cite{ref:wlee:1}.
As a consequence, we expect that, at ${\cal O}(a^2) \approx {\cal
O}(p^2)$, the pion spectra can be classified into 5 irreducible
representations of SO(4) taste symmetry: $\{\gamma_5 \otimes I\}$, $\{
\gamma_5 \otimes \xi_5\}$, $\{\gamma_5 \otimes \xi_\mu\}$, $\{
\gamma_5 \otimes \xi_{\mu5}\}$, $\{\gamma_5 \otimes \xi_{\mu\nu}\}$.
The pion with $\xi_5$ taste is the Goldstone mode corresponding to a
conserved axial symmetry which is broken spontaneously.

The splittings between pion multiplets reflect the taste symmetry
breaking directly.
As we improve the staggered fermion action and operators more, the
splittings get smaller and the taste symmetry gets closer to SU(4).
In other words, the splittings between pion multiplets can be used as
a non-perturbative probe to measure how much of the taste symmetry is
restored.

There have been a number of proposals to improve staggered fermions:
Fat7 \cite{ref:kostas:1}, AsqTad \cite{ref:kostas:1}, HYP
\cite{ref:anna:1}, $\overline{\rm Fat7}$ \cite{ref:wlee:2}, and
others.
Unimproved staggered fermions have three serious problems: (1) large
scaling violation, (2) large one-loop perturbative correction, and (3)
large taste symmetry breaking, which are well established through
a number of elaborate numerical studies.
It turns out that HYP and $\overline{\rm Fat7}$ improvement schemes
have the smallest one loop corrections ($\ll 5$ \%) to bilinear
operators \cite{ref:wlee:3}, which resolves the problem of large
perturbative corrections completely.
The AsqTad and HYP improvement schemes are already being used
extensively for numerical study of various physical observables, and
it is important to determine which scheme is the best.
Even though the perturbative comparison in Ref.~\cite{ref:wlee:3}
provides a hint that HYP and $\overline{\rm Fat7}$ schemes are better
than the AsqTad scheme, it was not conclusive at that time due to a
lack of non-perturbative evidence of improvement for the remaining
two issues.
In this context, it is quite important to compare the size of
splittings between pion multiplets for various improvement schemes,
which will tell us which improvement scheme is best to restore
the SU(4) taste symmetry.
We will address the issue of large scaling violation later when we
talk about $B_K$.
Here, let us focus on the issue of the taste symmetry breaking by
looking into splittings among pion multiplets.

\begin{figure}[t!]
\includegraphics[width=0.5\textwidth]{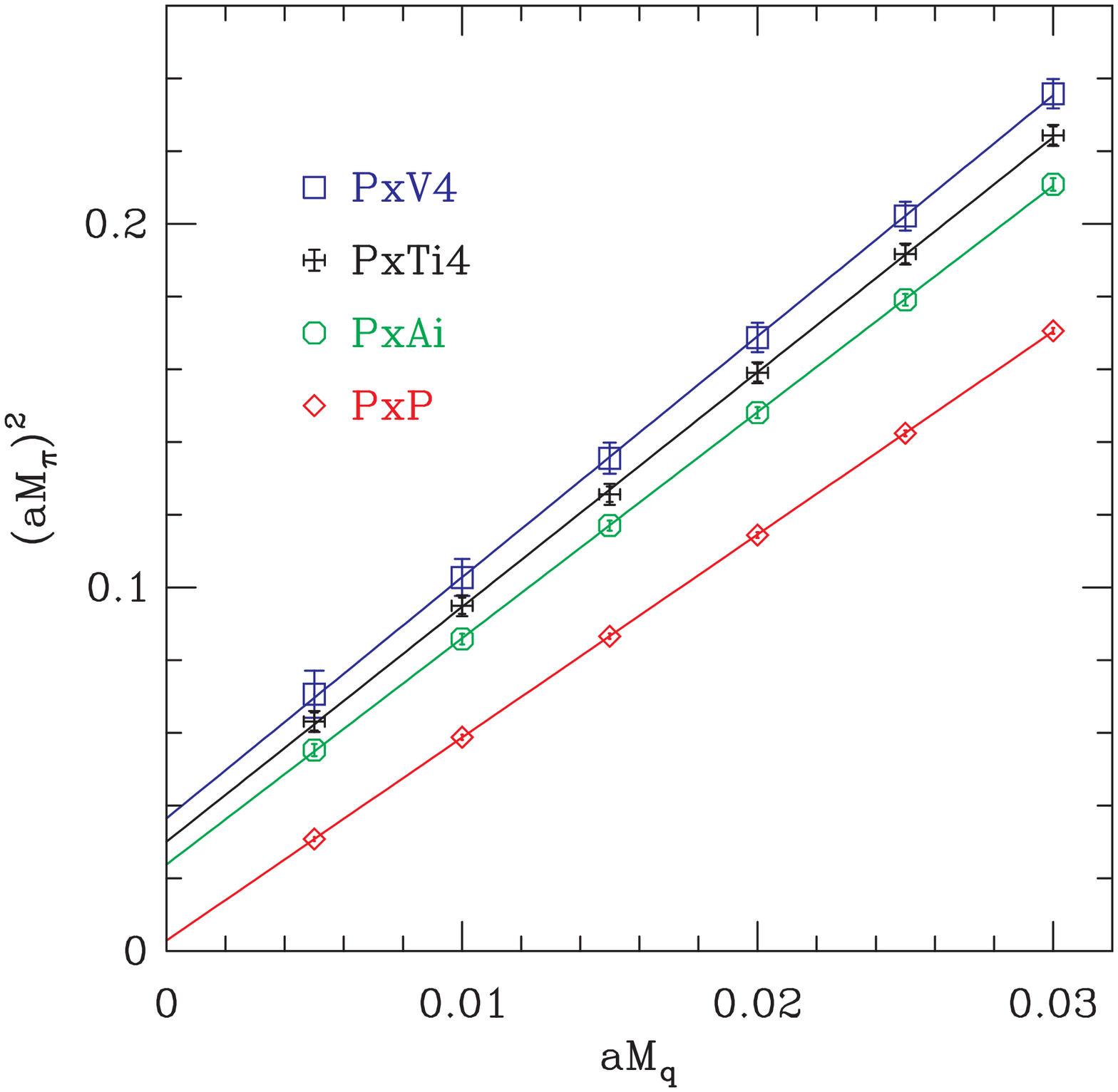}
\includegraphics[width=0.5\textwidth]{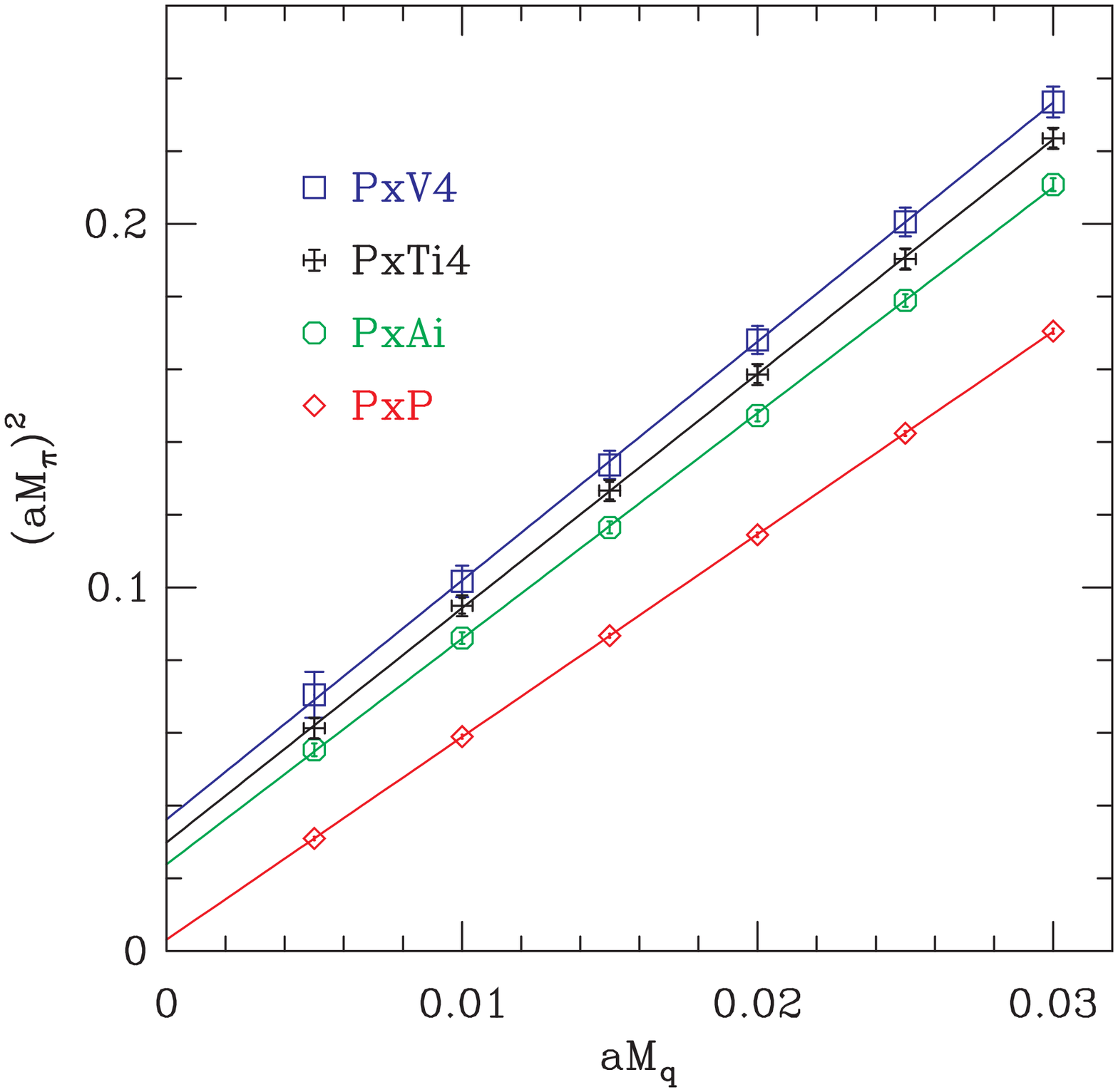}
\caption{$(a m_\pi)^2$ vs.~$am_q$ for unimproved staggered fermions:
  (left) using the Kluberg-Stern method;
  (right) using the Golterman method.}
\label{fig:mpisq:unimp}
\end{figure}

In order to probe the size of the taste symmetry breaking for
unimproved staggered fermions, we have calculated the pion multiplet
spectrum in quenched QCD ($\beta=6.0$ and $16^3\times 64$ lattice) and
preliminary results are presented in Ref.~\cite{ref:wlee:4}.
In Fig.~\ref{fig:mpisq:unimp}, we show $(am_\pi)^2$ as a function of
quark mass $am_q$ for various pion multiplets.
As discussed in Ref.~\cite{ref:wlee:4}, there are two independent
methods to construct the bilinear operators using staggered fermions:
(1) the Kluberg-Stern method and (2) the Golterman method.
In Fig.~\ref{fig:mpisq:unimp}, the left (right) plot is obtained using
the Kluberg-Stern (Golterman) method.
Note that there is essentially no difference between the Kluberg-Stern
and Golterman methods.
We also observe that the splittings among pion multiplets are
comparable to the light pion masses, which implies that ${\cal O}(a^2)
\approx {\cal O}(p^2)$.
In addition, note that the slopes are noticeably different for various
tastes, which implies that ${\cal O}(a^2 p^2)$ terms are so large as to
be noticeable in the case of unimproved staggered fermions, even
though they are of higher order in staggered chiral perturbation
theory.

\begin{figure}[t!]
\includegraphics[width=0.5\textwidth]{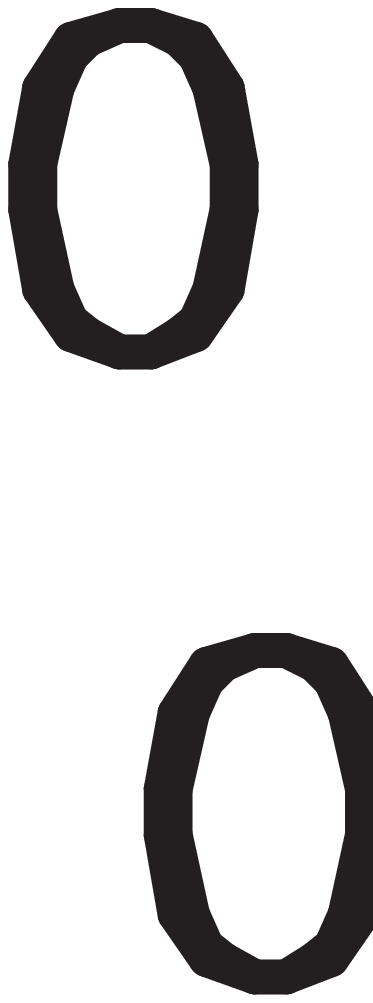}
\includegraphics[width=0.5\textwidth]{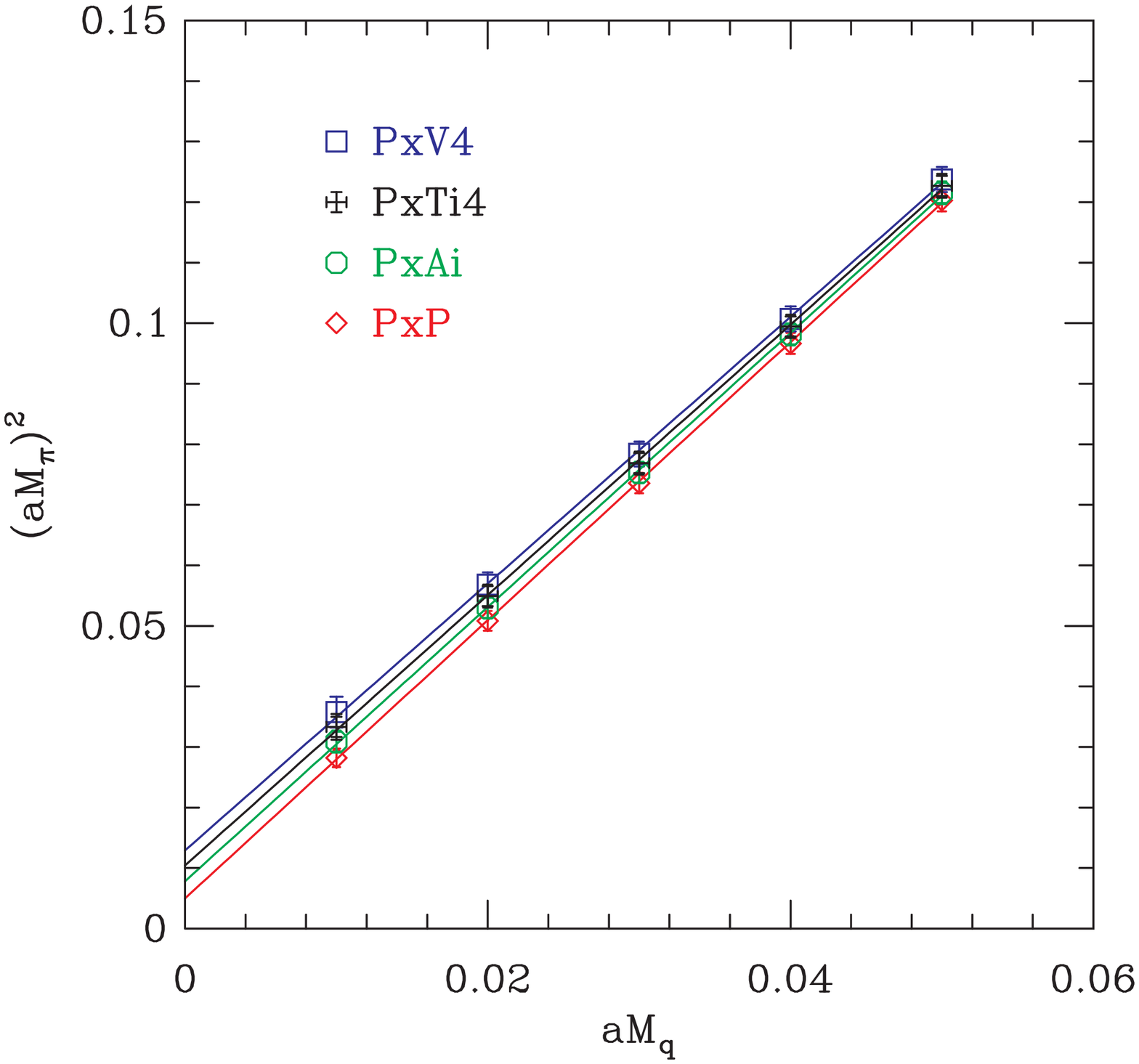}
\caption{$(a m_\pi)^2$ vs.~$am_q$:
  (left) for AsqTad staggered fermions;
  (right) for HYP staggered fermions (using the Golterman method).}
\label{fig:mpisq:hyp}
\end{figure}

In Fig.~\ref{fig:mpisq:hyp}, we compare the results of AsqTad
staggered fermions with those of HYP staggered fermions.
The AsqTad staggered fermion data comes originally from
Ref.~\cite{ref:milc:1} whereas the HYP staggered fermion data comes
from Ref.~\cite{ref:wlee:4}.
Since AsqTad staggered fermions use a Symanzik improved gauge action
with $a=0.12$fm and HYP staggered fermions use the unimproved Wilson
plaquette action with $a=0.1$fm, we expect that the scaling violations
associated with the background gauge configurations are similar in
both cases.
In the case of AsqTad staggered fermions, the size of the taste
symmetry breaking effect is comparable to the light pion masses, which
implies that ${\cal O}(a^2) \approx {\cal O}(p^2)$.
In addition, note that the slopes of pion multiplets with different
tastes are the same, which implies that the ${\cal O}(a^2 p^2)$ terms
are negligibly small for AsqTad staggered fermions.

In the right-hand side of Fig.~\ref{fig:mpisq:hyp}, we show data for
HYP staggered fermions.
In this plot, we observe two things. 
First, the splitting among pion multiplets are extremely suppressed
within $2\sigma$ of the data point error bar, which implies that
${\cal O}(a^2) \ll {\cal O}(p^2)$.
Second, we notice that the slopes of pion multiplets with different
tastes coincide within statistical uncertainty, which implies that
the ${\cal O}(a^2 p^2)$ terms are so small that the effect is not
detected at all.

Therefore, from Fig.~\ref{fig:mpisq:hyp}, we conclude that HYP
staggered fermions are remarkably better than AsqTad staggered
fermions with regards to reducing the taste symmetry breaking effect.
From Fig.~\ref{fig:mpisq:unimp} and Fig.~\ref{fig:mpisq:hyp}, the
${\cal O}(a^2 p^2)$ terms are so large for unimproved staggered fermions
that the effect was noticeable in the data, whereas these higher order
terms are negligibly small in the case of AsqTad and HYP staggered
fermions, which we can count as a big improvement.
Therefore, we can rank the improvement efficiency of various staggered
fermions with regards to taste symmetry restoration as follows:
\begin{equation}
\mbox{unimproved stag} < \mbox{AsqTad stag} < \mbox{HYP stag}
\end{equation}
More details on the comparison of various improvement schemes can be found
in Ref.~\cite{ref:wlee:4}.

\section{$f_\pi$ and $f_K$}
\label{sec:fk}
A precise determination of low energy constants in QCD is one of the
major goals that we want to achieve on the lattice.
In particular, $f_\pi$ and $f_K$ are relatively cheap to calculate on
the lattice and their analysis is straight forward.
The real challenge is that we need to produce a large ensemble of
gauge configurations which includes sea quark back-reaction (sea quark
determinant) with $N_F = 2+1$ flavors ($m_u = m_d \ne m_s$).
This procedure is quite expensive computationally.
Then, we compute valence quark propagators on the gluon backgrounds to
calculate hadron spectrum and hadronic matrix elements.
To get as much physics information as possible from the valuable gauge
configurations with specific sea quark masses, we use many values of
valence quark masses, {\em i.e.} not just the same values as the sea
quark masses.
This physical set-up is known as ``partially quenched QCD''.
It has been studied with a number of lattice fermion formulations: (1)
AsqTad staggered fermions, (2) Wilson clover fermions, (3) Wilson
twisted mass fermions, (4) domain wall fermions, and (5) overlap
fermions.
In the following, we discuss the cases of AsqTad staggered fermions,
domain wall fermions, and a mixed action approach (valence quarks are
domain wall fermions and sea quarks are AsqTad staggered fermions).

\subsection{$f_\pi$ and $f_K$ using AsqTad staggered fermions}
\label{subsec:fk:stag}
Staggered fermions are fastest to simulate on the lattice, and have an
exact axial symmetry which protects the quark mass from additive
renormalization and makes it possible to calculate weak matrix
elements reliably.
The physical quark flavors are each represented by a staggered fermion
field, so for three quark flavors there is an exact SU(3) flavor
symmetry (when the quark masses are degenerate) as in the continuum,
along with SU(4) taste symmetry of each of the staggered fermion
fields (which is broken at finite lattice spacing but restored in the
continuum limit).
The hadronic operators can be taken to have a specific taste, so the
lattice formulation with staggered fermions reproduces the continuum
one at tree level in chiral perturbation theory.
However, pions with all kinds of tastes appear in loops, and through
this taste violations enter everywhere.
Hence, it is necessary to take into account taste violations (directly
related to discretization effects) within chiral perturbation theory
in order to fit the data and extract physical information with high
precision.
A systematic tool to incorporate taste violations into the data
analysis is the staggered chiral perturbation theory
\cite{ref:wlee:1,ref:bernard:1}.

Another caveat regarding broken taste symmetry is that we take
$\sqrt[4]{{\rm Det}(D+m)}$ as the fermion determinant to get a single
taste per sea quark flavor.
This is often called the ``fourth-root trick''.
There is a possibility that $\sqrt[4]{{\rm Det}(D+m)}$ produces
non-perturbatively violations of locality and universality in the
continuum limit, which might cause staggered fermions not to reproduce
QCD.
However, recent arguments and numerical investigations give encouraging
indications that this possibility does not occur.
For more details on this issue, see Ref.~\cite{ref:sharpe:1}.
\begin{figure}[t!]
\includegraphics[width=0.5\textwidth]{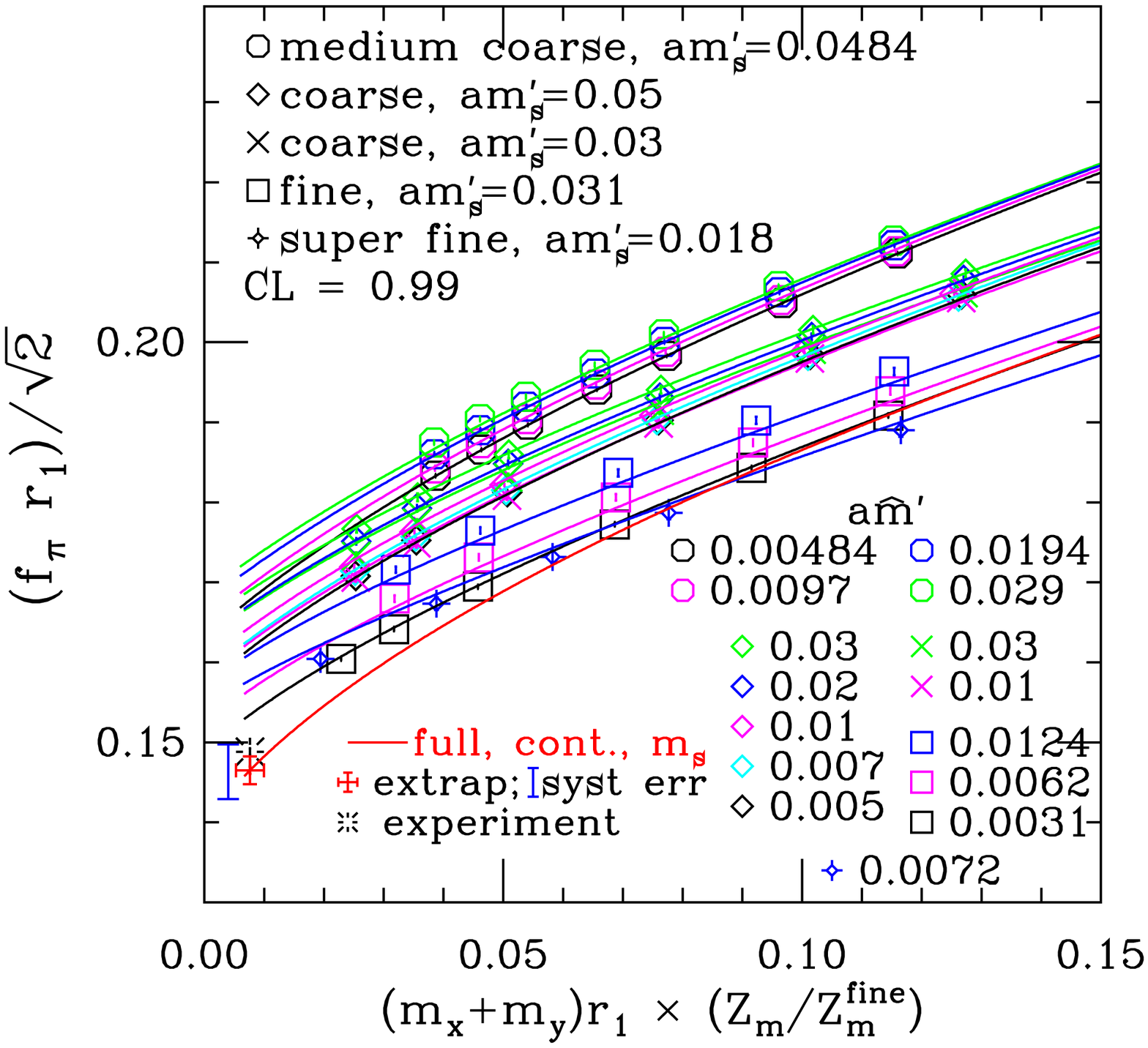}
\includegraphics[width=0.5\textwidth]{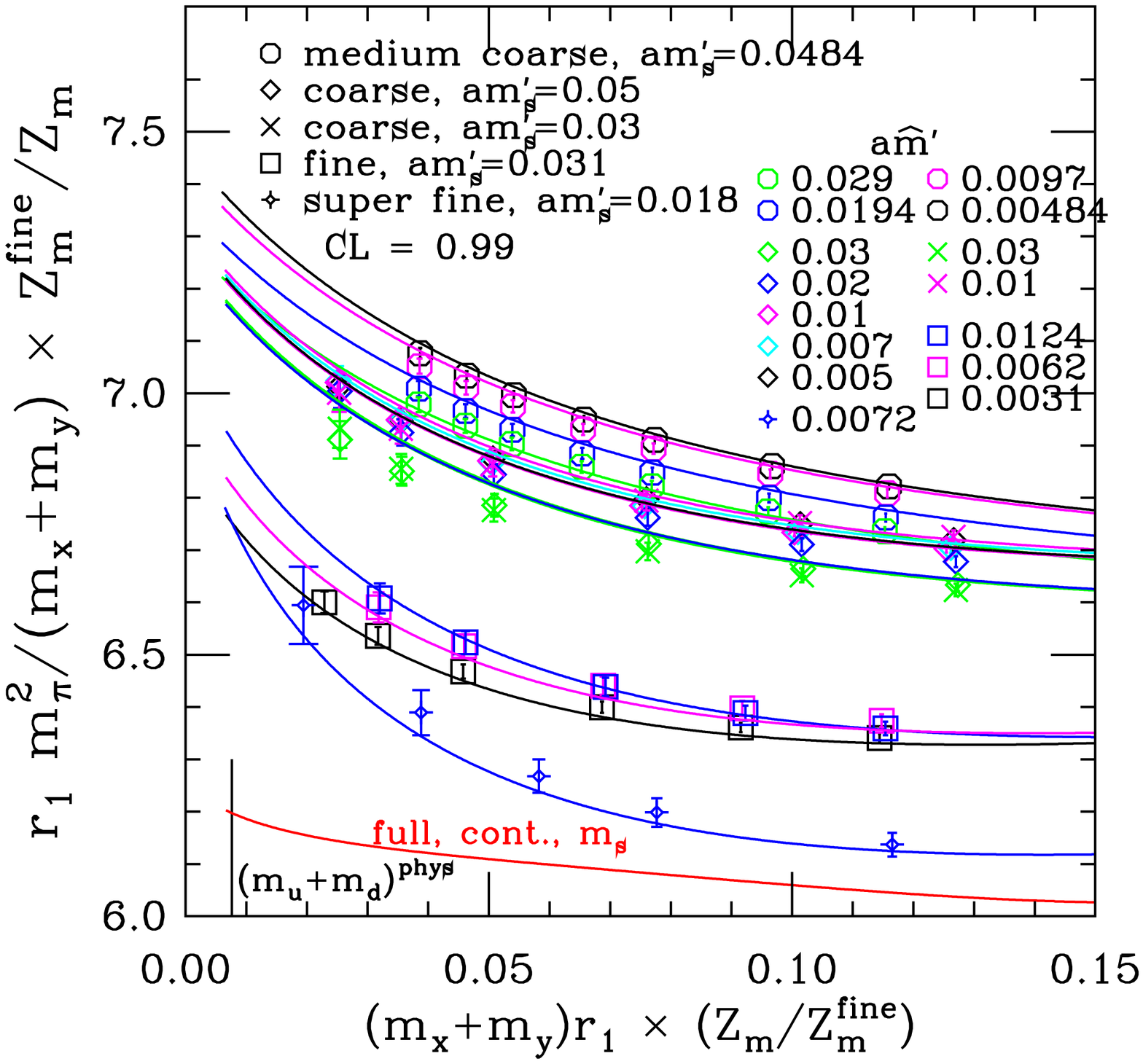}
\caption{$f_\pi$ vs.~$m_x + m_y$ (left) and $m_\pi^2 / (m_x + m_y)$
vs.~$m_x + m_y$ (right). The data are obtained using AsqTad
staggered fermions in $N_F = 2+1$ partially quenched QCD.}
\label{fig:fpi:milc}
\end{figure}
Recently, MILC has been generating dynamical gauge configurations
with $N_F=2+1$ flavor sea quarks (AsqTad staggered fermions).
These configurations include coarser lattices ($a=0.15$fm),
coarse lattices ($a=0.12$fm), fine lattices ($a=0.09$fm) and
super-fine lattices ($a=0.06$fm).
On these lattices, the strange sea quark mass ($M_s$) of simulation
runs the range of $0.7 m_s \lesssim M_s \lesssim 1.2 m_s$.
The light sea quark mass of simulation goes down to $m_u = m_d \approx
11$ MeV ($m_\pi \approx 240$ MeV).
The physical volumes of the lattices range from $(2.4 fm)^3$ to $(3.4
fm)^3$.
Using these gauge configurations, MILC has been calculating full pion
multiplet spectrum and decay constants for the Goldstone pions and
kaons.
The results are presented in Fig.~\ref{fig:fpi:milc}.
In this extensive calculation, MILC concentrates on the analysis of
masses and decay constants of Goldstone pions and kaons.
This analysis also involves the masses of non-Goldstone pions and
kaons; these enter in one-loop chiral logs at NLO in staggered chiral
perturbation theory \cite{ref:bernard:2}.
In the data analysis, it is necessary to cut out the data points
of heavier quark masses one by one in order to get good fits to
the form suggested by staggered chiral perturbation theory. 
The most updated results are summarized in Table \ref{tab:fpi:milc}.
For more details, see Ref.~\cite{ref:milc:2}.
\begin{table}[t!]
\begin{center}
\begin{tabular}{ c | c }
\hline
observable & value \\
\hline
$f_\pi$  & $128.6 \pm 0.4 \pm 3.0$ MeV \\
$f_K$    & $155.3 \pm 0.4 \pm 3.1$ MeV \\
$f_K/f_\pi$ & $1.208(2)_{-14}^{+7}$  \\
$V_{us}$ & $0.2223_{-14}^{+26}$  \\
$m_s(2 {\rm GeV},\overline{\rm MS})$ & 90(0)(5)(4)(0) MeV \\
$m_{u,d}(2 {\rm GeV},\overline{\rm MS})$ & 3.3(0)(2)(2)(0) MeV \\
$m_s/m_{u,d}$ & 27.2(0)(4)(0)(0) \\
$L_5$ & $2.0(3)(2) \times 10^{-3}$ \\
\hline
\end{tabular}
\end{center}
\caption{Results of data analysis by MILC.}
\label{tab:fpi:milc}
\end{table}

\subsection{$f_\pi$ and $f_K$ using domain wall fermions}
\label{subsec:fk:dwf}
\begin{figure}[t!]
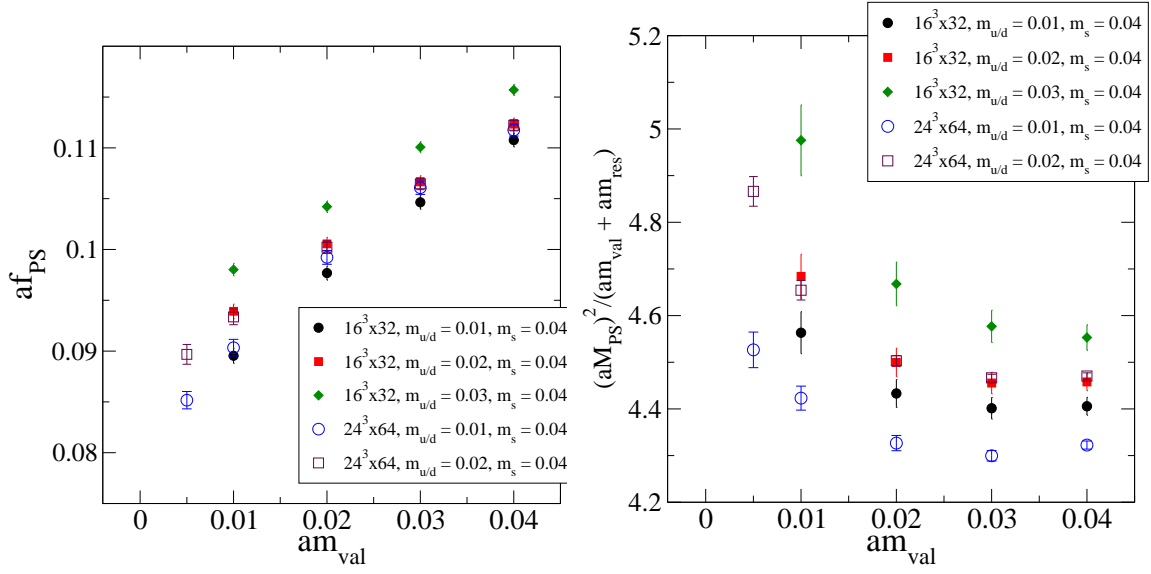

\includegraphics[width=0.5\textwidth]{fpi_RBC_lat06.eps}
\includegraphics[width=0.5\textwidth]{mpisq_RBC_lat06.eps}
\caption{$af_\pi$ vs.~$am_{\rm val}$ (left) and $(am_\pi)^2 / (m_{\rm
val} + m_{\rm res})$ vs.~$m_{\rm val}$ (right). The data were obtained
using domain wall fermions in $N_F = 2+1$ partially quenched QCD.}
\label{fig:fpi:rbc}
\end{figure}

In the domain wall fermion formulation, the breaking of chiral symmetry
is due only to the finite length of the 5th dimension, which allows a
small additive renormalization to quark masses.
We call this a ``residual quark mass''; it is typically only a few MeV
and hence under control.
Domain wall fermions are computationally typically two orders of
magnitude more expensive than improved staggered fermions.
Since we can use continuum chiral perturbation theory for the analysis
of domain wall fermion data, the theoretical interpretation is simpler
and more straightforward compared to improved staggered fermions.
\begin{table}[h!]
\begin{center}
\begin{tabular}{ c | c }
\hline
parameter & value \\
\hline
lattice  & $16^3 \times 32$ and $24^3 \times 64$ \\
$L_s$    & 16 \\
$aM_5$   & 1.8 \\
$a$      & 0.125 fm  \\
$am_s$   & 0.04  \\
$am_{u,d}$   & 0.01, 0.02, 0.03  \\
\hline
\end{tabular}
\end{center}
\caption{Parameters of dynamical domain wall fermion simulation by RBC
and UKQCD.}
\label{tab:fpi:rbc}
\end{table}
The RBC and UKQCD collaborations use domain wall fermions with Iwasaki
gluon action to generate dynamical gauge configurations with $N_F=2+1$
sea quark flavors.
The simulation parameters are summarized in Table~\ref{tab:fpi:rbc}.
These lattices are comparable with the coarse MILC lattices ($a=0.12$
fm).
These groups measured pion decay constants and pion spectra in
partially quenched QCD, and the preliminary results are shown in
Fig.~\ref{fig:fpi:rbc}.
They put elaborate effort into fitting the data of $f_\pi$ and
$m_\pi^2/m_q$ simultaneously to the NLO form suggested by partially
quenched chiral perturbation theory.
It turns out that the simultaneous fits fail to describe the data.
They believe that this failure is due to the fact that the quark
masses are too heavy to interpret the data based on chiral perturbation
theory.
For more details, see Ref.~\cite{ref:rbc:1}.

\subsection{$f_K/f_\pi$ using a mixed action scheme}
\label{subsec:fk:mixed}
Another possibility in partially quenched QCD is to use a mixed action
scheme.
This is the approach taken by the LHPC and NPLQCD collaborations
\cite{ref:savage:1}.
In this approach, the sea quarks are AsqTad staggered fermions and
the valence quarks are domain wall fermions.
Specifically, they do HYP smearing over the MILC gauge configurations
with $N_F=2+1$ AsqTad staggered sea quarks and calculate the valence
quark propagators using domain wall fermions over these configurations
to measure $f_K/f_\pi$.
In this approach, the valence quark masses are tuned such that the
pion masses match the Goldstone pion masses of AsqTad sea quarks.

\begin{figure}[t!]
\begin{center}
\includegraphics[width=0.7\textwidth]{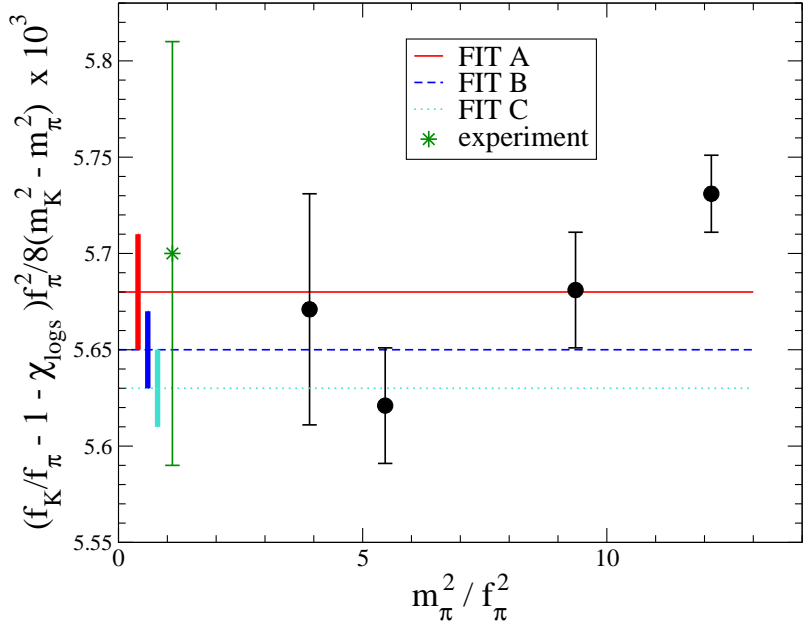}
\end{center}
\caption{$L_5(f^0_\pi)$ vs.~$m_\pi^2/f_\pi^2$ calculated by NPLQCD, using a
mixed action scheme.}
\label{fig:fpi:nplqcd}
\end{figure}

In the continuum SU(3) chiral perturbation theory, one can show that
$f_K/f_\pi$ is given, at the next-to-leading (NLO) order, by
\begin{eqnarray}
\frac{f_K}{f_\pi} &=& 1 + \chi_{\log} + 8 y L_5(\mu) \\
y &=& \frac{m_K^2 - m_\pi^2}{f_\pi^2} \\
L_5(f^0_\pi) &=& L_5(\mu) - \frac{3}{8} \frac{1}{16\pi^2} \log(f^0_\pi/\mu) 
\end{eqnarray}
where $\mu$ is a renormalization scale in chiral perturbation theory,
$\chi_{\log}$ represents all the NLO chiral log corrections
collectively, and $f^0_\pi$ is the physical value of pion decay
constant.
In Fig.~\ref{fig:fpi:nplqcd}, we show $L_5(f^0_\pi)$ calculated using
the mixed action scheme.
It is interesting to note that the final value of $f_K/f_\pi$ obtained
by NPLQCD is consistent with MILC, even though they apply the
continuum chiral perturbation theory to analyze data of the mixed
action scheme.
For more details, see Ref.~\cite{ref:savage:1}.

\section{Indirect CP violation and $B_K$}
\label{sec:bk}
In nature, there are two kinds of CP violations in the neutral kaon
system: indirect CP violation and direct CP violation.
The neutral kaon states in nature are described as CP eigenstates
with a tiny impurity which is parametrized as $\epsilon$.
In experiment, $\epsilon$ is determined precisely as follows:
\begin{equation}
\epsilon = (2.280 \pm 0.013) \times 10^{-3} \times \exp(i\pi/4)
\end{equation}
From the standard model, it is possible to express $\epsilon$ in terms of
$K^0-\bar{K}^0$ mixing as follows \cite{ref:buras:1}:
\begin{eqnarray}
  \epsilon &=& C_\epsilon \ \exp(i\pi/4) \ {\rm Im}\lambda_t
  \ X \ \hat{B}_K  \\
  X &=& {\rm Re} \lambda_c [ \eta_1 S_0(x_c) - \eta_3 S_3(x_c,x_t) ]
  - {\rm Re} \lambda_t \eta_2 S_0(x_t) \\
  \lambda_i &=& V_{is}^* V_{id}, \qquad x_i = m_i^2 / M_W^2 \\
  C_\epsilon &=& \frac{G_F^2 f_K^2 m_K M_W^2}{6 \sqrt{2} \pi^2 \Delta m_K}
\end{eqnarray}
Here, note that $S_i$ (the Inami-Nam functions) and $\eta_i$ are
determined reliably from perturbation theory; they represent the
remaining effect of heavy particles when they are integrated out.
The kaon bag parameter $\hat{B}_K$ is defined as
\begin{eqnarray}
  B_K &=& \frac{\langle \bar{K}_0 | [\bar{s} \gamma_\mu (1-\gamma_5) d]
    [\bar{s} \gamma_\mu (1-\gamma_5) d] | K_0 \rangle }{
    \frac{8}{3} \langle \bar{K}_0 | \bar{s} \gamma_\mu\gamma_5 d | 0 \rangle
    \langle 0 | \bar{s} \gamma_\mu\gamma_5 d | K_0 \rangle }
  \\
  \hat{B}_K &=& C(\mu) B_K(\mu)
  \\
  C(\mu) &=& \alpha_s(\mu)^{-\frac{\gamma_0}{2 \beta_0}}
  [ 1 + \alpha_s(\mu) J_3 ]
\end{eqnarray}
where $\mu$ is a renormalization scale of the $\Delta S = 2$ operator,
and $\hat{B}_K$ is renormalization group invariant.
$V_{ij}$ is the CKM matrix element of the standard model.
The theoretical challenge is to calculate $B_K$ with such a high
precision that it over-constrains the unitarity of the CKM matrix.

\begin{figure}[t!]
\includegraphics[width=0.6\textwidth]{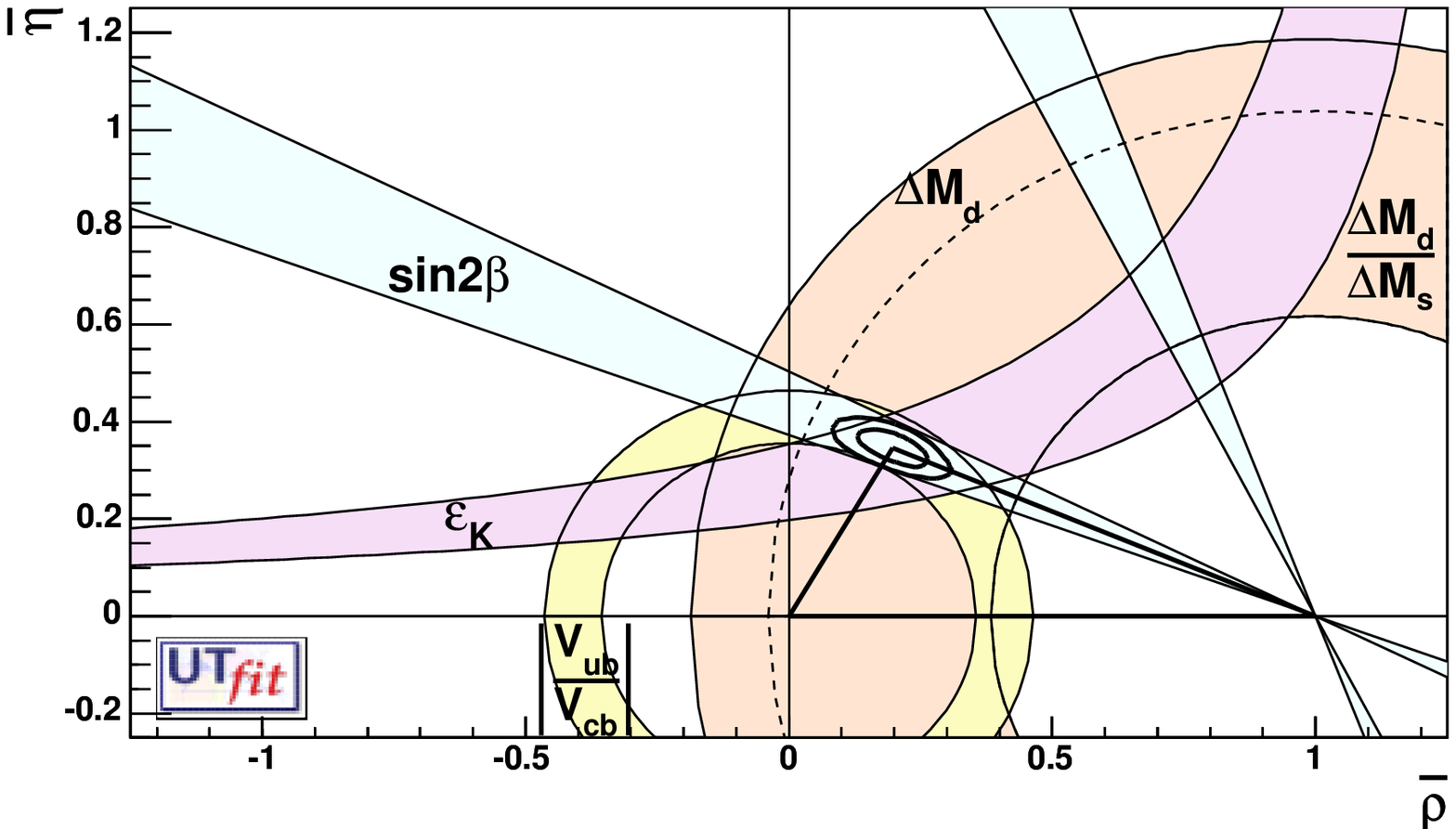}
\includegraphics[width=0.4\textwidth]{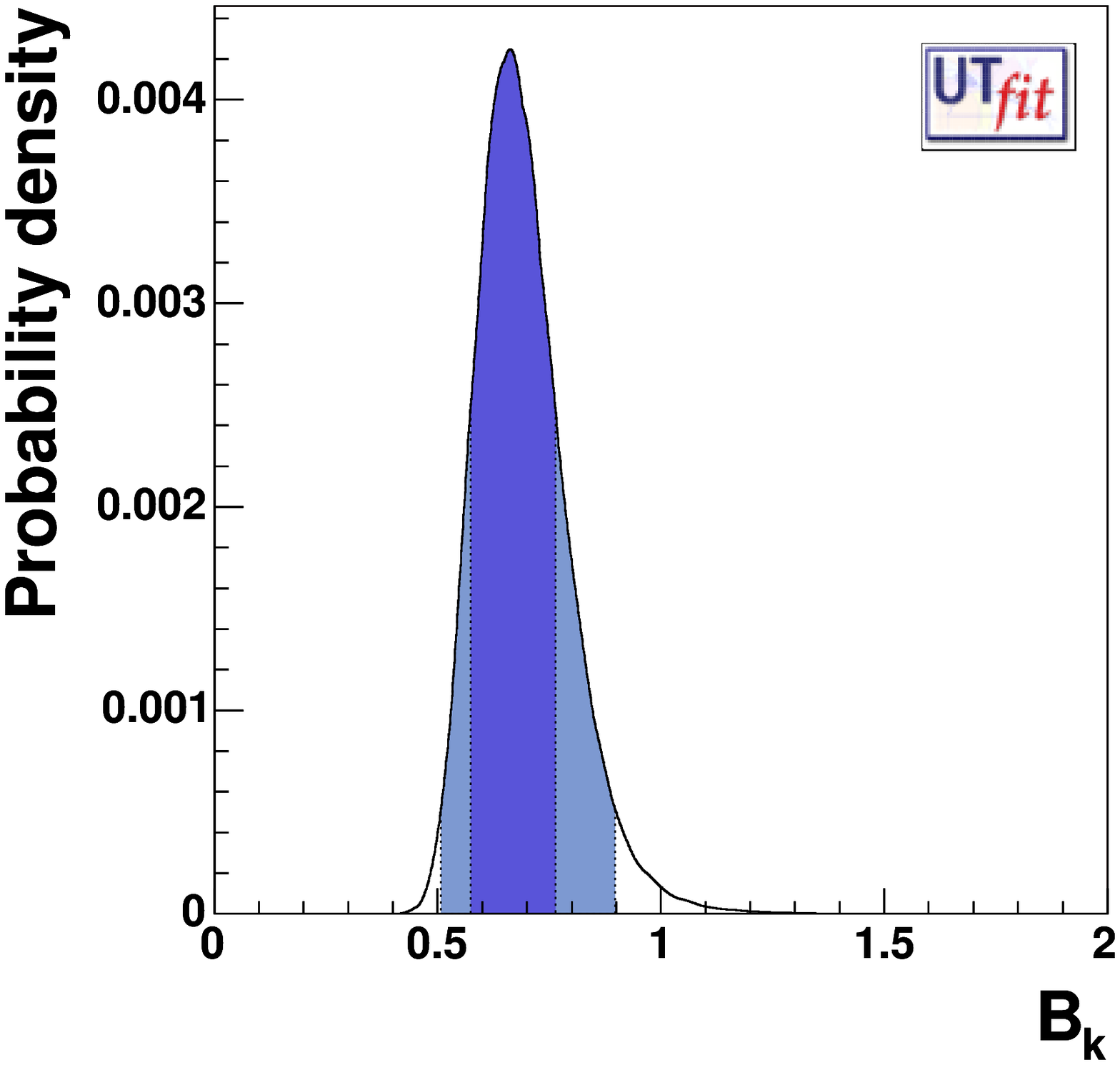}
\caption{$\hat{B}_K$ determined from the unitarity triangle.}
\label{fig:bk:ckm}
\end{figure}

It is possible to determine $\hat{B}_K$ from the untarity triangle if
we obtain the CKM matrix elements from other experimental results.
In Fig.~\ref{fig:bk:ckm}, we show the result of $\hat{B}_K$ determined
from the unitarity triangle.
The result reported in CKM 2005 is $\hat{B}_K = 0.68 \pm 0.10$.

\subsection{$B_K$ in quenched QCD}
\label{subsec:bk:quenched}
\begin{figure}[t!]
\begin{center}
\includegraphics[width=0.7\textwidth]{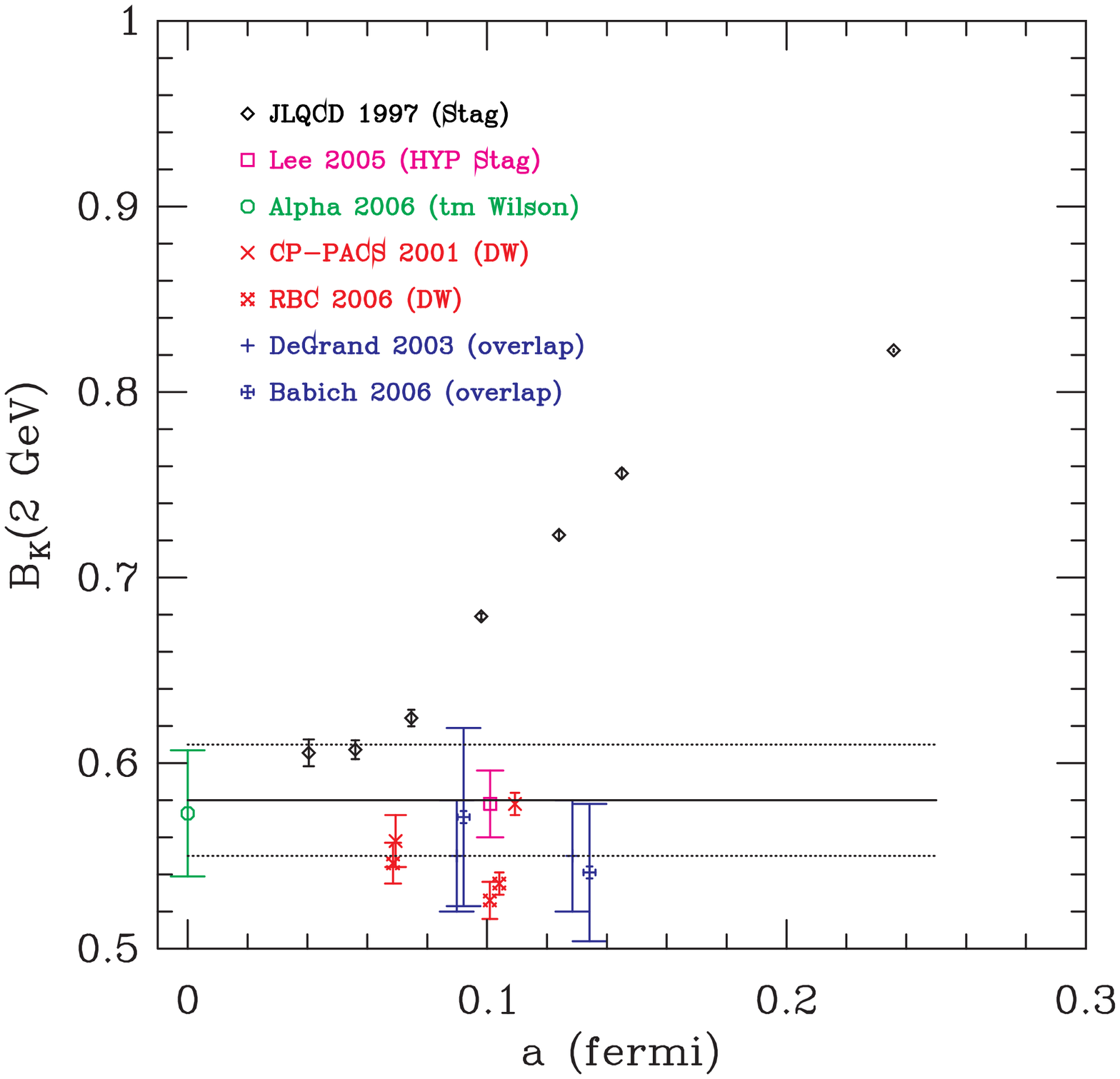}
\end{center}
\caption{$B_K(2GeV,{\rm NDR})$ calculated in quenched QCD.}
\label{fig:bk:quenched}
\end{figure}

There have been a number of efforts to calculate $B_K$ in quenched QCD
using staggered fermions \cite{ref:jlqcd:1,ref:wlee:5}, twisted mass
(TM) Wilson fermions \cite{ref:alpha:1}, domain wall fermions
\cite{ref:cp-pacs:1,ref:rbc:2}, and overlap fermions
\cite{ref:degrand:1,ref:lellouch:1}.
In Fig.~\ref{fig:bk:quenched}, the results of these calculations are
summarized.
We observe a large scaling violation in the calculation of $B_K$
using unimproved staggered fermions by JLQCD \cite{ref:jlqcd:1}.
However, it is possible to reduce the scaling violation down to a
negligibly small level using the HYP improvement scheme to improve the
staggered fermion action and operators \cite{ref:wlee:5}.
We will address the issue of scaling violation in detail in
Sec.~\ref{subsec:bk:scale}.
An interesting demonstration of the progress in $B_K$ calculations is
that the value of $B_K$ calculated using the TM Wilson fermions ends
up being in agreement with other $B_K$ calculations using different
fermion formulations \cite{ref:alpha:1}.
The calculation of $B_K$ with TM Wilson fermions has been discussed in
great detail in the plenary talk by Pena \cite{ref:alpha:2} so we skip
all the details here.
The values of $B_K$ calculated by CP-PACS \cite{ref:cp-pacs:1} and by
RBC \cite{ref:rbc:2} using domain wall fermions are in good agreement with
those of other fermion formulations.
The $B_K$ values calculated by two independent groups
\cite{ref:degrand:1,ref:lellouch:1} using overlap fermions are also in
agreement with those of other fermion formulations, although they have
relatively large error bars.
In these calculations, it is assumed that $m_s = m_d$, {\em i.e.} the
effects of non-degenerate quark masses are neglected.
A simple-minded world average estimate of quenched $B_K$ with
degenerate quarks is $B_K(2{\rm GeV}, {\rm NDR}) = 0.58(4)$.

\subsection{Scaling violation in $B_K$ and its reduction 
with improved staggered fermions}
\label{subsec:bk:scale}
\begin{figure}[t!]
\begin{center}
\includegraphics[width=0.7\textwidth]{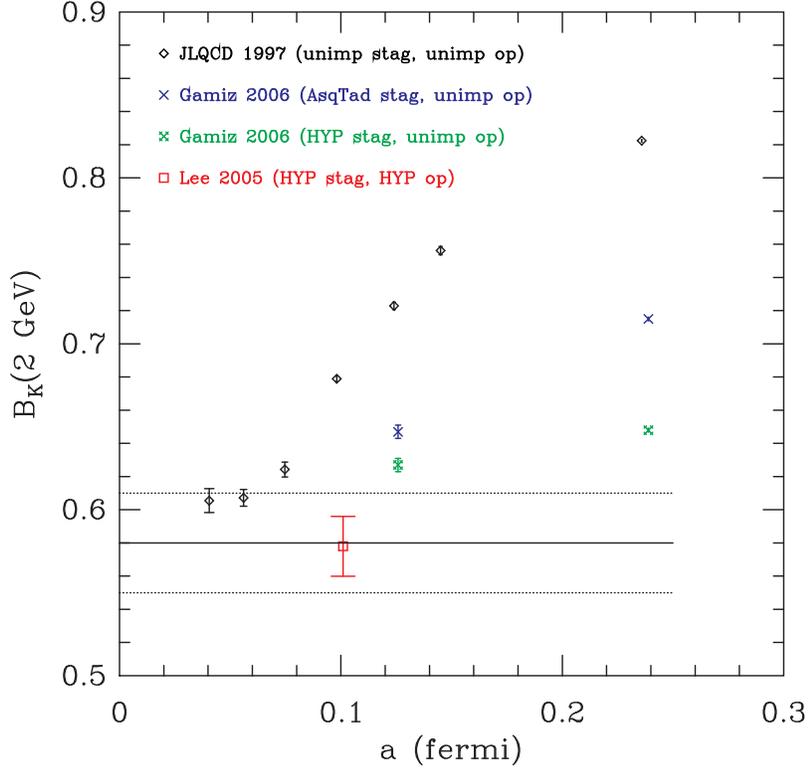}
\end{center}
\caption{$B_K(2GeV,{\rm NDR})$ calculated using various staggered
fermion formulations.}
\label{fig:bk:sv:stag}
\end{figure}

We now focus on the scaling violation in $B_K$ calculated using
staggered fermions.
In Fig.~\ref{fig:bk:sv:stag}, we show $B_K$ data as a function of
lattice spacing $a$ (fm).
Here, we observe that there is a large scaling violation (=
discretization error) in the data calculated by JLQCD using the
unimproved staggered fermion action and unimproved operators
(constructed using thin links).
The blue (green) data points in Fig.~\ref{fig:bk:sv:stag} represent
the $B_K$ values calculated using the AsqTad (HYP) staggered fermion
action and unimproved operators \cite{ref:gamiz:1}.
We notice that in this case the scaling violations are reduced by
about 50\%.
This comparison tells us that the HYP staggered fermion action is much
more efficient in reducing the scaling violations than the AsqTad
staggered fermion action.
However, we also observe that the remaining scaling violations are still
substantial.
The red data points in Fig.~\ref{fig:bk:sv:stag} represents the $B_K$
values obtained using the HYP staggered fermion action and the
improved operators with HYP fat links \cite{ref:wlee:5}.
In this case, the $B_K$ value at $a=0.1$ fm is consistent with the
world average which is obtained by extrapolation to the continuum
limit of ($a=0$).
This implies that, once we improve the staggered fermion action and
operators with HYP fat links, the scaling violations in $B_K$ are
reduced down to a negligibly small level.

From Fig.~\ref{fig:bk:sv:stag}, we conclude that we can rank the
improvement efficiency of various staggered fermions with regards to
reducing the scaling violations as follows:
\begin{equation}
\mbox{unimproved stag} < \mbox{AsqTad stag} < \mbox{HYP stag}
\end{equation}
Combining this conclusion with that of Sec.~\ref{sec:spec}, we see
that HYP staggered fermions are remarkably better than AsqTad
staggered fermions in reducing the taste symmetry breaking effect, the
scaling violations, and the one-loop perturbative corrections
simultaneously \cite{ref:wlee:3}.

\subsection{$B_K$ in unquenched QCD with $N_F=2+1$}
\label{subsec:bk:full}
\begin{figure}[t!]
\includegraphics[width=0.5\textwidth]{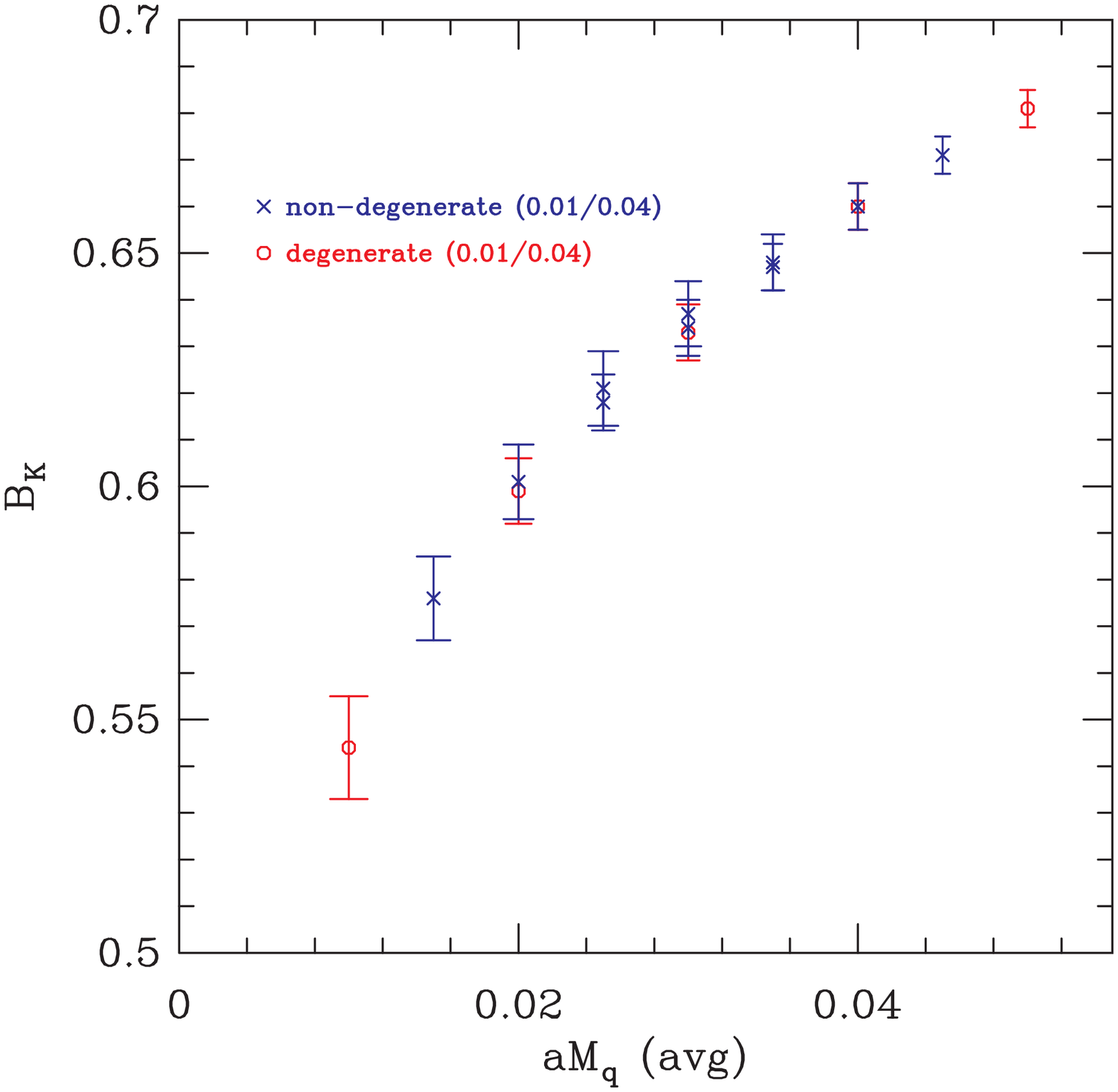}
\includegraphics[width=0.5\textwidth]{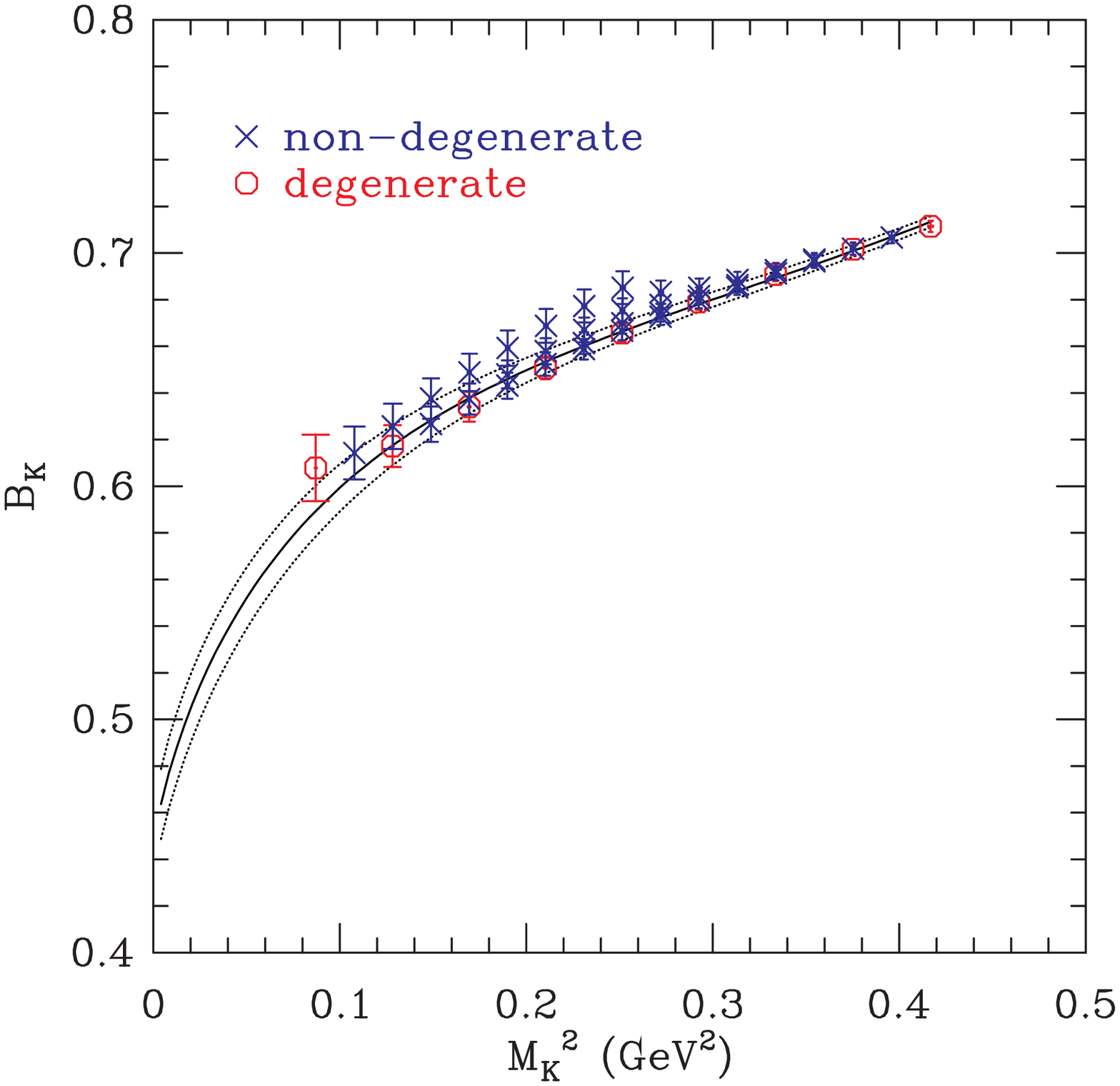}
\caption{$B_K^{\rm bare}$ in unquenched QCD with $N_F=2+1$; (left)
calculated by RBC and UKQCD collaborations using domain wall fermions
and (right) calculated by Lee {\em et al.} using a mixed action scheme
in which valence quarks are HYP staggered fermions and sea quarks are
AsqTad staggered fermions.}
\label{fig:bk:rbc:wlee}
\end{figure}

Recently, there have been two serious attempts to calculate $B_K$ in
unquenched QCD with $N_F=2+1$.
In Fig.~\ref{fig:bk:rbc:wlee}, we show $B_K$ data in $N_F=2+1$ flavor
QCD. 
On the left-hand side of Fig.~\ref{fig:bk:rbc:wlee}, we show the $B_K$
data of the RBC and UKQCD collaborations calculated using domain wall
fermions at $a=0.125$ fm \cite{ref:rbc:bk}.
On the right-hand side of Fig.~\ref{fig:bk:rbc:wlee}, we present the
$B_K$ data ($a=0.12$ fm) obtained using a mixed action scheme in which
valence quarks are HYP staggered fermions and sea quarks are AsqTad
staggered fermions \cite{ref:wlee:6}.
The $B_K$ data of domain wall fermions does not have a noticeable
dependence on the non-degenerate quark masses.
However, we observe that the $B_K$ data of improved staggered fermions
shows a noticeable dependence on the non-degenerate quark masses,
corresponding to about $3\sigma$ difference between the degenerate
and non-degenerate quark mass combinations near the physical kaon
mass.
Certainly, this is a puzzle.
One possible explanation is that there is a significant difference in
statistics between the two sets of data: 75 measurements for the
domain wall fermion data and 640 measurements for the improved
staggered fermion data.
In other words, it might be that this dependence on the non-degenerate
masses only shows up with extremely high statistics.
At any rate, this issue needs further investigation in the future.

There has been another attempt to calculate $B_K$ in unquenched QCD
using the AsqTad staggered fermions and unimproved operators, where
$B_K$ is calculated with degenerate quark masses set to $am_s/2$
\cite{ref:gamiz:1}.
Since this work does not address the main issue of this section,
we omit the details.

\section{Direct CP violation and ${\rm Re}(\epsilon'/\epsilon)$}
\label{sec:eprime}
Direct CP violation occurs in kaon decays via electroweak interactions
where the CP symmetry is explicitly broken.
This process is parametrized by $\epsilon'/\epsilon$.
Recently the KTeV and NA48 collaborations have determined 
${\rm Re}(\epsilon'/\epsilon)$ from high statistics experiments
\cite{ref:ktev:1,ref:na48:1}.
The results are
\begin{eqnarray}
{\rm Re}(\epsilon'/\epsilon) &=& (20.7 \pm 2.8) \times 10^{-4}
\qquad \mbox{(KTeV)}  \\
&=& (14.7 \pm 2.2) \times 10^{-4} \qquad \mbox{(NA48)}
\end{eqnarray}
Here, note that both KTeV and NA48 claim that ${\rm
Re}(\epsilon'/\epsilon)$ is large and positive ($\gg 1.0 \times
10^{-4}$).

From the standard model, we can derive the theoretical expression for
$\epsilon'/\epsilon$ as follows \cite{ref:buras:1}:
\begin{eqnarray}
  \epsilon'/\epsilon &=& {\rm Im}(V_{ts}^* V_{td})
  \Big[ P^{(1/2)} - P^{(3/2)} \Big] \exp(i \Phi)
  \label{eq:eprime} \\
  P^{(1/2)} &=& r \sum_{i=3}^{10} y_i(\mu)
  \langle Q_i(\mu) \rangle_0
  (1 - \Omega_{\rm IB} ) \\
  P^{(3/2)} &=& \frac{r}{\omega} \sum_{i=7}^{10} y_i(\mu)
  \langle Q_i(\mu) \rangle_2 \\
  \langle Q_i \rangle_I &=&
  \langle \pi\pi(I)| Q_i | K \rangle \\
  r &=& \frac{ G_F \omega }{ 2 | \epsilon | {\rm Re}(A_0) }
\end{eqnarray}
where $\mu$ is a renormalization scale of the $Q_i$ operators.
Here, $P^{(1/2)}$ and $P^{(3/2)}$ represent the $\Delta I = 1/2$ and
$\Delta I = 3/2$ contributions to $\epsilon'/\epsilon$ respectively.
The Wilson coefficient functions $y_i(\mu)$ can be determined reliably
from the perturbation theory in a specific renormalization scheme.
Note that the phase shift $\Phi = \Phi_{\epsilon'} - \Phi_\epsilon
\approx 0$ and $\omega = 0.045$ (the $\Delta I = 1/2$ rule).
The parameter $\Omega_{\rm IB}$ represents the isospin breaking effect.

As can be seen in Eq.~\ref{eq:eprime}, $\epsilon'/\epsilon$ is
obtained as a result of destructive interferences between $P^{(1/2)}$
and $P^{(3/2)}$.
In particular, $P^{(3/2)}$ is fully dominated by electroweak penguin
contributions, especially $Q_8$.
$P^{(1/2)}$ on the other hand is governed by QCD penguin
contributions, especially $Q_6$.
The main sources of uncertainty in the calculation of
$\epsilon'/\epsilon$ are the hadronic matrix elements $\langle
\pi\pi(I)| Q_i | K \rangle$.
Hence, the theoretical challenge on the lattice is to calculate
$\langle \pi\pi(I)| Q_i | K \rangle$ reliably.

The RBC and CP-PACS collaborations reported that $\epsilon'/\epsilon$
is small and negative.
Their calculation is done in quenched QCD with operators defined in
the SU(3) flavor group theory using domain wall fermions
\cite{ref:rbc:e'/e,ref:cp-pacs:e'/e}.
However, Golterman and Pallante have pointed out that there is an
serious ambiguity in the lattice version of left-right QCD penguin
operators ($Q_6$ and $Q_5$) because the flavor group in quenched QCD
is not SU(3) but SU(3|3) \cite{ref:golterman:Q6}.
It turns out that the ambiguity in $Q_6$ has such a large effect on
$\epsilon'/\epsilon$ that it can flip the sign of $\epsilon'/\epsilon$
in quenched QCD: this has been demonstrated in calculations using HYP
staggered fermions \cite{ref:wlee:7,ref:wlee:8}.
Recently, the RBC Collaboration have investigated this issue in great
detail \cite{ref:rbc:Q6}, which will be discussed in the next section.
In addition, Golterman and Pallante point out that there is another
ambiguity in $Q_i$ ($i=1,2,3,4$) in quenched QCD, which has a serious
effect on ${\rm Re}(A_0)$ and the sub-leading contribution of $Q_4$
on $\epsilon'/\epsilon$ \cite{ref:golterman:Q4}.
In other words, it is not possible to calculate reliably even the
sub-leading contribution of $Q_4$ as well as the leading contribution
of $Q_6$ to $\epsilon'/\epsilon$ in quenched QCD.
We will address this issue later.

\subsection{Ambiguity in $Q_6$ and $Q_5$}
\label{sec:Q6}
We now focus on the left-right QCD penguin operators, especially $Q_6$
which makes the dominant contribution to $P^{(1/2)}$.
In the standard model, $Q_6$ is defined as
\begin{equation}
Q_6 = \bar{s} \gamma_\mu (1 - \gamma_5) d 
\sum_{q} \bar{q} \gamma_\mu (1 + \gamma_5) q
\end{equation}
$Q_6$ belong to the (8,1) irreducible representation of
the $SU(3)_L \otimes SU(3)_R$ flavor symmetry group.
Unfortunately, $Q_6$ mixes with $\tilde{Q}_i$ ($i=3,4,5,6$) in
(partially) quenched QCD, where for example, $\tilde{Q}_6$ is defined
as
\begin{equation}
\tilde{Q}_6 =\bar{s} \gamma_\mu (1 - \gamma_5) d
\sum_{\tilde{q}} \bar{\tilde{q}} \gamma_\mu (1 + \gamma_5) \tilde{q}
\end{equation}
where $\tilde{q}$ represents a ghost quark.
Here, note that the flavor symmetry of (partially) quenched QCD is not
$SU(3)$ but $SU(3+N_F|3)$ where $N_F$ is the number of sea quark
flavors.
In quenched QCD ($N_F=0$), Golterman and Pallante propose to
decompose $Q_6$ as follows:
\begin{eqnarray}
Q_6 &=& Q_6^{QS} + Q_6^{QNS} \\
Q_6^{QS} &=& \frac{1}{2}\bar{s} \gamma_\mu (1 - \gamma_5) d
\sum_{Q} \bar{Q} \gamma_\mu (1 + \gamma_5) Q \\
Q_6^{QNS} &=& \frac{1}{2} \bar{s} \gamma_\mu (1 - \gamma_5) d
\sum_{Q} \bar{Q} \gamma_\mu (1 + \gamma_5) A Q \\
\end{eqnarray}
where $Q = (u,d,s,\tilde{u},\tilde{d},\tilde{s})$ and
$A = {\rm diag}(1,1,1,-1,-1,-1)$ is a diagonal matrix.
See Ref.~\cite{ref:golterman:Q6,ref:rbc:Q6} for the details of the
group theoretical interpretation on this decomposition.
$Q_6^{QS}$ can be represented by a set of operators in quenched chiral
perturbation theory which have analogues in continuum chiral
perturbation theory.
However, $Q_6^{QNS}$ introduces a new low energy constant which
is called $\alpha_q^{NS}$.
In addition, Golterman and Peris have predicted that $\alpha_q^{NS}
\approx 60 \alpha_{q1}^{(8,1)}$ \cite{ref:golterman:a_NS}, which
implies that the matrix elements of $Q_6$ are completely dominated by
unphysical leading contribution from the lattice artifact
$\alpha_q^{NS}$.
Unfortunately, the work by the RBC and CP-PACS collaborations in
Ref.~\cite{ref:rbc:e'/e,ref:cp-pacs:e'/e} contains this artifact
as it is.
As mentioned, the work of Ref.~\cite{ref:wlee:7,ref:wlee:8} shows
clearly that this ambiguity is so large that it can even flip the sign
of $\epsilon'/\epsilon$.
\begin{figure}[t!]
\begin{center}
\includegraphics[width=0.7\textwidth]{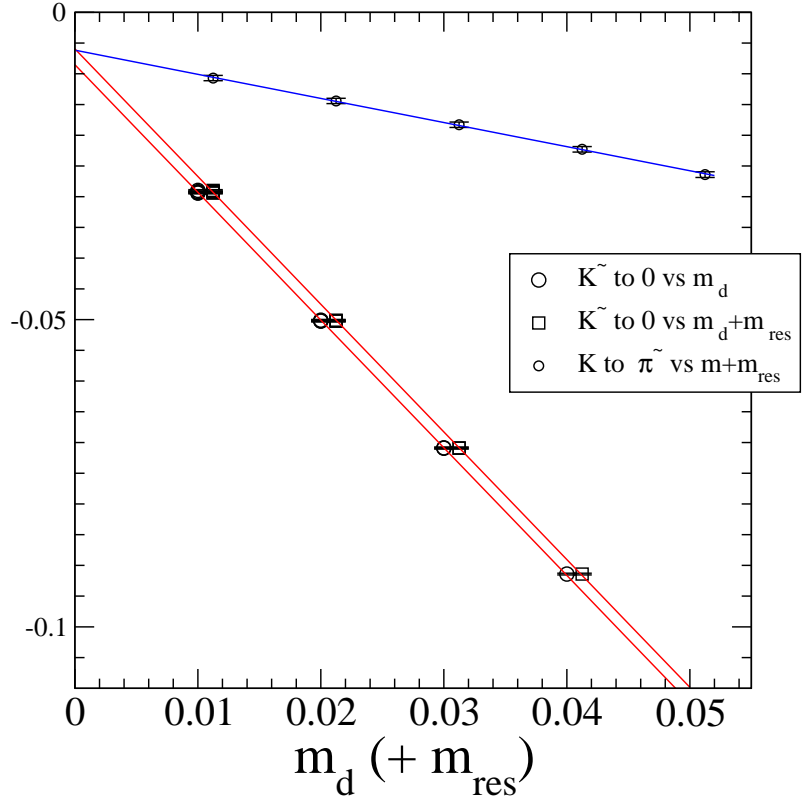}
\end{center}
\caption{$\alpha_q^{NS}$ in quenched QCD calculated 
by the RBC collaboration.}
\label{fig:a_NS:rbc}
\end{figure}
Recently, the RBC collaboration has measured the low energy constant
$\alpha_q^{NS}$ by two independent methods \cite{ref:rbc:Q6}: (1)
using the hadron matrix element of $\tilde{K} \rightarrow 0$ which is
originally suggested by Golterman and Pallante, and (2) using the $K^+
\rightarrow \tilde{\pi}^+$ matrix element.
In the case of domain wall fermions, the 2nd method has the advantage
that one does not have to worry about the residual quark mass
$m_{res}$, because the divergent NLO term vanishes in the limit of
degenerate quark masses.
In Fig.~\ref{fig:a_NS:rbc}, we show the procedure to obtain
$\alpha_q^{NS}$ by extrapolation to the chiral limit.
The figure demonstrates that both methods produce the same result in
the chiral limit and it is consistent with the prediction by Golterman
and Peris.
See Ref.~\cite{ref:rbc:Q6} for more details.

\subsection{Ambiguity in $Q_3$ and $Q_4$}
\label{sec:Q4}
Here, we focus on the left-left QCD penguin operators ($Q_3$, $Q_4$,
part of $Q_1$ and $Q_2$, $\cdots$).
As an example, let us consider $Q_3$ and $Q_4$:
\begin{eqnarray}
Q_3 &=& \bar{s}_\alpha \gamma_\mu (1 - \gamma_5) d_\alpha
\sum_{q} \bar{q}_\beta \gamma_\mu (1 - \gamma_5) q_\beta
= (\bar{s}_\alpha d_\alpha)_L \sum_{q} (\bar{q}_\beta q_\beta)_L \\
Q_4 &=& \bar{s}_\alpha \gamma_\mu (1 - \gamma_5) d_\beta
\sum_{q} \bar{q}_\beta \gamma_\mu (1 - \gamma_5) q_\alpha
= (\bar{s}_\alpha d_\beta)_L \sum_{q} (\bar{q}_\beta q_\alpha)_L
\end{eqnarray}
Both $Q_3$ and $Q_4$ belong to the (8,1) irreducible representation of
the $SU(3)_L \otimes SU(3)_R$ flavor symmetry group, which also
corresponds to $I=1/2$.
For the following discussion it is convenient to change the operator
basis as follows:
\begin{equation}
Q_\pm = Q_3 \pm Q_4
\end{equation}
Here, note that $Q_-$ ($Q_+$) is symmetric (anti-symmetric) in
the two anti-quark flavor indices and two quark flavor indices.
Once more, the problem is that the flavor symmetry group of
(partially) quenched QCD is not $SU(3)$ but $SU(3+N_F|3)$.
Hence, $Q_\pm$ is reducible under $SU(3+N_F|3)$.
We can decompose $Q_\pm$ as follows:
\begin{eqnarray}
Q_\pm &=& \frac{3}{N_F} Q^{S}_\pm + Q^{A}_\pm \\
Q^{S}_\pm &=& 
(\bar{q}_\alpha \Lambda q_\alpha)_L (\bar{q}_\beta q_\beta)_L
\pm (\bar{q}_\alpha \Lambda q_\beta)_L (\bar{q}_\beta q_\alpha)_L 
\label{eq:Q4:singlet}\\
Q^{A}_\pm &=&
(\bar{q}_\alpha \Lambda q_\alpha)_L (\bar{q}_\beta A q_\beta)_L
\pm (\bar{q}_\alpha \Lambda q_\beta)_L (\bar{q}_\beta A q_\alpha)_L
\end{eqnarray}
where $q$ represents (valence plus sea) quark fields as well as ghost
quark fields.
The spurion fields $\Lambda$ and $A$ are defined as
\begin{eqnarray}
& & \Lambda_{ij} = \delta_{i3} \delta_{j2} \\
& & A = {\rm diag} (1 - \frac{3}{N_F}, 1 - \frac{3}{N_F}, 1 - \frac{3}{N_F},
-\frac{3}{N_F}, \cdots ,-\frac{3}{N_F},)
\end{eqnarray}
In this decomposition, the $Q^{S}_\pm$ part produces those terms which
have analogues in continuum SU(3) chiral perturbation theory.
However, the $Q^{A}_\pm$ part introduces new low energy constants which
are pure lattice artifacts.
In quenched QCD, the decomposition is slightly different; see
Ref.~\cite{ref:golterman:Q4} for the details.
At any rate, this unphysical lattice artifact from the $Q^{A}_\pm$
part appears at the leading order in (partially) quenched chiral
perturbation theory.
This has a serious effect on ${\rm Re}(A_0)$ through $Q_1$ and $Q_2$
as well as $\epsilon'/\epsilon$ through $Q_4$.
Consequently, it is not possible to calculate reliably not only the
leading contribution of $Q_6$ but also the sub-leading contribution of
$Q_4$ to $\epsilon'/\epsilon$ in quenched QCD.
Moreover, the analysis of ${\rm Re}(A_0)$ given in
Ref.~\cite{ref:rbc:e'/e,ref:cp-pacs:e'/e,ref:wlee:9} did not take into
account the effect of this lattice artifact from the $Q^{A}_\pm$ part,
which implies that the results are contaminated by this as well.

\subsection{Solution to the problem}
\label{sec:sol}
In the previous sections \ref{sec:Q6} and \ref{sec:Q4}, we explained
that it is not possible to reliably calculate $\epsilon'/\epsilon$ in
quenched QCD due to a fundamental ambiguity in $Q_6$ and $Q_4$.
Is there any way to get around the problem?
One possible solution is the following 
\begin{itemize}
\item Do the numerical study in partially quenched QCD with $N_F=3$ or
$N_F=2+1$. This insures that the low energy constants are physical.
\item Calculate the hadronic matrix elements of $Q_6$ and $Q_5$ using
an singlet operator of $SU(6|3)_R$ to remove the unphysical lattice
artifact.
\item Calculate the hadronic matrix elements of $Q_4$ and $Q_3$ using
linear combinations of the $Q^{S}_\pm$ operators defined in
Eq.~\ref{eq:Q4:singlet}.
\end{itemize}
In this way, the calculations will extract only those low energy
constants which are physically relevant to the direct CP violation
without contamination by non-physical lattice artifacts.

\section{$K\rightarrow\pi\pi$ decay}
\label{sec:k-pipi}
There have been two independent approaches to calculating the
non-leptonic kaon decay process $K \rightarrow \pi\pi$ on the lattice:
a direct approach and an indirect one.
In the indirect approach, we calculate the hadronic matrix elements of
$K\rightarrow \pi$ and $K \rightarrow 0$ and reconstruct $K
\rightarrow \pi\pi$ amplitudes out of them using the chiral perturbation
theory.
Since it is relatively easy and computationally cheap, the indirect
method has been widely used to calculate the $K\rightarrow\pi\pi$
amplitudes \cite{ref:rbc:e'/e,ref:cp-pacs:e'/e,ref:wlee:7}.
In the direct approach, we calculate the $K\rightarrow \pi\pi$ matrix
elements directly with the final state pions carrying a physical
momentum of $|\vec{p}| = 206$ MeV.
This method has a number of difficulties on the lattice.
In Ref.~\cite{ref:testa:1}, Maiani and Testa pointed out a ``no-go
theorem'': we can not obtain $K \rightarrow \pi(\vec{p})
\pi(-\vec{p})$ but only $K \rightarrow \pi(\vec{0}) \pi(-\vec{0})$ on
the lattice.
This is seen as follows. 
We want to obtain the matrix element of $\langle \pi(\vec{p})
\pi(-\vec{p}) | Q | K \rangle$ by calculating the following four-point
function $G$:
\begin{equation}
G(t,t_\pi, t_K) = 
\langle 0 | \phi_\pi(\vec{p},t_\pi) \phi_\pi(-\vec{p},t_\pi) 
Q(t) \phi_K(t_K) | 0 \rangle
\end{equation}
However, the asymptotic behavior of $G$ is
\begin{equation}
\lim_{t_K \ll t \ll t_\pi } G(t,t_\pi, t_K) =
Z_{\vec{p}\vec{0}} \cdot \exp(-E_{\pi\pi,{\vec{0}}}(t_\pi-t))
\cdot \langle \pi(\vec{0}) \pi(\vec{0}) | Q | K \rangle 
\cdot Z_K \cdot \exp(-E_K(t-t_K)) 
\end{equation}
where
\begin{eqnarray}
Z_K &=& \langle K | \phi_K | 0 \rangle
\\
Z_{\vec{p}\vec{q}} &=& 
\langle 0 | \phi_\pi(\vec{p}) \phi_\pi(-\vec{p}) | 
\pi(\vec{q}) \pi(-\vec{q}) \rangle 
\\
E_{\pi\pi,{\vec{p}}} &=& 2 \sqrt{m_\pi^2 + \vec{p}^2} 
\end{eqnarray}
Hence, we need $\langle \pi(\vec{p}) \pi(-\vec{p}) | Q | K \rangle$,
but we can get only the information on $\langle \pi(\vec{0})
\pi(\vec{0}) | Q | K \rangle$, because $ | \pi(\vec{0}) \pi(\vec{0})
\rangle$ is the ground state of two pions with periodic boundary
condition in the spatial direction.

\subsection{Circumventing the Maiani-Testa theorem} 
\label{sec:MT:bypass}
Is there any way to get around the Maiani-Testa theorem?
Luscher and Wolff proposed the diagonalization method in
Ref.~\cite{ref:luscher:1}. 
In this method, we calculate the following correlation function:
\begin{eqnarray}
C_{nm}(t) &=& \langle \Omega_n (t) \Omega_m(0) \rangle
\\
\Omega_n(t) &=& \phi_\pi(\vec{p}_n, t) \phi_\pi(-\vec{p}_n, t)
\end{eqnarray}
Then, we can construct the following matrix $M(t,t_0)$:
\begin{equation}
M(t,t_0) = C(t_0)^{-1/2} \cdot C(t) \cdot C(t_0)^{-1/2} 
\end{equation}
By diagonalizing the $M(t,t_0)$ matrix, we can obtain the
two pion eigenstates and eigenvalues.
Of course, this takes enormous amount of computing resources because
we need to take a large number of momenta in order to get a few low
eigenvalues, and the signal gets poor as we insert the pion momentum.
A number of lattice groups applied this method to the $I=2$ elastic
pion scattering problem \cite{ref:fiebig:1,ref:cp-pacs:phase-shift}.

One way to get around the Maiani-Testa theorem is to modify the
boundary condition (BC) for the pions \cite{ref:kim:1}.
Of course, there are many different choices to modify BC: (1) H parity
BC, (2) G parity BC, (3) twisted BC and so on.
As an example, let us consider the H parity BC.
\begin{eqnarray}
& & H(u,d) = (-u,d)
\\
& & H | \pi^{\pm} \rangle = - | \pi^{\pm} \rangle
\\
& & H | \pi^0 \rangle = + | \pi^0 \rangle
\end{eqnarray}
The H parity BC causes the ground state of $\pi^{\pm}$ to have a
finite momentum $|\vec{p}| = 1/2 (2\pi/L)$.
Hence, if we construct two pion state out of $\pi^+$ and $\pi^-$ with
the H parity BC, we can calculate $\langle \pi(\vec{p}) \pi(-\vec{p}) | Q
| K \rangle$.

Another way to get around the Maiani-Testa theorem is to use the
moving frame (LAB) instead of the center of mass (CM) frame.
In Ref.~\cite{ref:gottlieb:1,ref:christ:1,ref:sachrajda:1}, it was
proposed to put a non-zero total momentum into the final two pion state
and initial kaon state.
In other words, calculate $\langle \pi(\vec{P}) \pi(\vec{0}) | Q |
K(\vec{P}) \rangle$.
Of course, we need a formula which can convert this result on the LAB
frame into that on the CM frame.

\subsection{Lellouch-Luscher formula in the moving frame}
\label{sec:ll:LAB}
Recently, Lellouch and Luscher have provided a formula in the CM frame
which connects the $K\rightarrow \pi\pi$ amplitude calculated in a
finite box (on the lattice) with the $K\rightarrow \pi\pi$ amplitude
in the infinite volume \cite{ref:luscher:2}.
Let us define $A$ as the on-shell decay amplitude in the infinite volume
of the CM frame and $M$ as that in the finite volume of the CM frame.
Their relation is
\begin{equation}
|A|^2 = 8 \pi \bigg( \frac{ E_{\pi\pi} }{ p } \bigg)^3
\bigg\{ p \frac{\partial \delta(p)} {\partial p}
+ q \frac{\partial \phi(q)} {\partial q} \bigg\}
|M|^2
\label{eq:ll:CM}
\end{equation}
where $\delta$ is the scattering phase shift and
is obtained from the following relationship:
\begin{eqnarray}
& & E_{\pi\pi} = 2 \sqrt{m_\pi^2+p^2} = m_K
\\
& & q = \frac{L}{2\pi}\cdot p
\\
& & \delta(p) = n\pi - \phi(q)
\\
& & \tan(\phi(q)) = - \frac{q\pi^{3/2}}{Z_{00}(1;q^2)}
\\
& & Z_{00}(1;q^2) = \frac{1}{\sqrt{4\pi}} \sum_{n \in Z^3}
\frac{1}{n^2 - q^2}
\end{eqnarray}
However, this formula is valid only in the CM frame.
Hence, we need a similar relationship which works in the LAB frame.
The latter is given in
Ref.~\cite{ref:gottlieb:1,ref:christ:1,ref:sachrajda:1}.
We can use the Lellouch-Luscher formula with the following
modifications for $K(P) \rightarrow \pi(P)\pi(0)$:
\begin{eqnarray}
& & (E^{LAB}_{\pi\pi})^2 = E^2_{\pi\pi} + P^2
\\
& & |M_{LAB}| = \frac{1}{\gamma} |M_{CM}|
\\
& & \gamma = \frac{E^{LAB}_{\pi\pi}}{E_{\pi\pi}}
\end{eqnarray}
The scattering phase shift is modified as follows:
\begin{eqnarray}
& & E_{\pi\pi} = 2 \sqrt{m_\pi^2+p^2} = m_K
\\
& & q = \frac{L}{2\pi}\cdot p
\\
& & \delta(p) = n\pi - \phi_{\vec{P}}(q)
\\
& & \tan(\phi_{\vec{P}}(q)) = 
- \frac{\gamma q\pi^{3/2}}{Z^{\vec{P}}_{00}(1;q^2)}
\\
& & Z^{\vec{P}}_{00}(1;q^2) = 
\frac{1}{\sqrt{4\pi}} \sum_{n \in Z^3}
\frac{1}{n_1^2 + n_2^2 + \gamma^{-2} (n_3+ LP/4\pi)^2 - q^2}
\end{eqnarray}
where we assume that $\vec{P} = (0,0,2\pi/L)$.
The relation between $|A|$ and $|M_{LAB}|$ is
\begin{equation}
|A|^2 = 8 \pi \gamma^2 \bigg( \frac{ E_{\pi\pi} }{ p } \bigg)^3
\bigg\{ p \frac{\partial \delta(p)} {\partial p}
+ q \frac{\partial \phi_{\vec{P}}(q)} {\partial q} \bigg\}
|M_{LAB}|^2
\label{eq:ll:LAB}
\end{equation}
Note that Eq.~\ref{eq:ll:LAB} becomes identical to Eq.~\ref{eq:ll:CM}
in the limit of $\vec{P} = 0$ and $\gamma =1$.

\subsection{Numerical study in the moving frame}
\label{sec:num:LAB}
\begin{figure}[t!]
\centering \includegraphics[height=1.0\textwidth,angle=270]{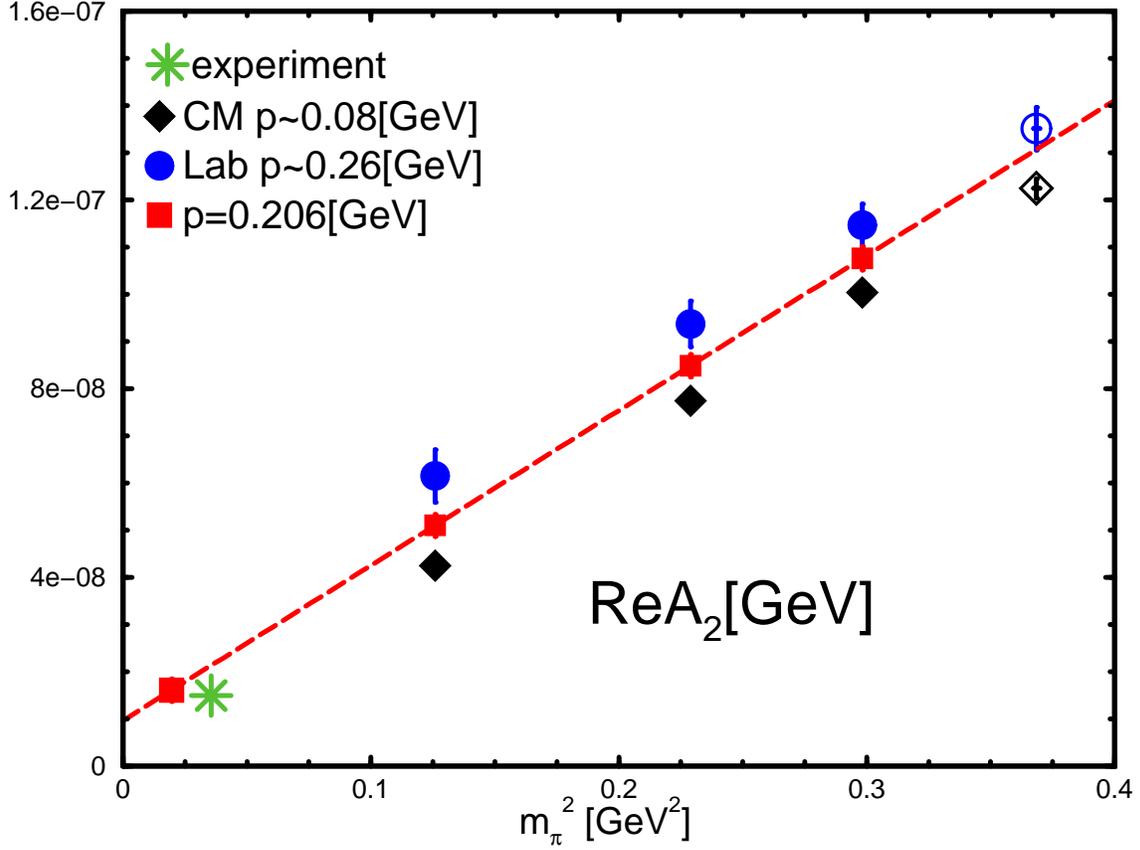}
\caption{${\rm Re}(A_2)$ vs.~$m_\pi^2$ in quenched QCD calculated
  using domain wall fermions.}
\label{fig:ReA2}
\end{figure}
Recently, the RBC collaboration has calculated the $K(P) \rightarrow
\pi(P)\pi(0)$ amplitudes in the LAB frame in quenched QCD
\cite{ref:rbc:LAB}.
This numerical study is performed at $1/a=1.3$ GeV on the $16^3\times
32$ lattice with $L_s = 12$ using domain wall fermions and the DBW2
gauge action.
Using the formula of Eq.~\ref{eq:ll:LAB}, the result in the LAB frame 
is converted into that of the infinite volume in the CM frame.
A preliminary result for ${\rm Re}(A_2)$ is shown in
Fig.~\ref{fig:ReA2}.
%

%
% EDIT
%

\section{$K_{l3}$ decay}
\label{sec:kl3}
The $K_{l3}$ decays are the kaon beta decay processes: $K^+\rightarrow
\pi^0 l^+ \nu_l$ and $K^0 \rightarrow \pi^- l^+ \nu_l$, where $l$
represents the electron or the muon.
These decay channels play an important role in determining the
CKM matrix element $|V_{us}|$.
The decay rate of the $K_{l3}$ processes is given
in Ref.~\cite{ref:leutwyler:1}:
\begin{equation}
\Gamma = \frac{G_F^2}{192\pi^3} m_K^5 \cdot |V_{us}|^2 \cdot C^2 
\cdot |f_+(0)|^2 \cdot I \cdot (1+\delta)
\end{equation}
where $C$ is a Clebsch-Gordon coefficient (1 for $K^0$ decay;
$1/\sqrt{2}$ for $K^+$ decay), $I$ represents the phase space
integral, and $\delta$ represents radiative corrections from
electroweak and electromagnetic interactions.
Here, $f_+(0)$ is a form factor which is defined for the neutral kaon
decay as
\begin{equation}
\langle \pi(p') | \bar{s} \gamma_\mu u | K(p) \rangle
= (p_\mu + p'_\mu) f_+(q^2) + q_\mu f_-(q^2)
\end{equation}
where $q = p - p'$ represents the momentum transfer.
For later discussion, it is convenient to define the scalar form factor
$f_0(q^2)$ as
\begin{eqnarray}
f_0(q^2) &=& f_+(q^2) + \frac{q^2}{m_K^2 - m_\pi^2} f_-(q^2) \\
\xi(q^2) &=& \frac{f_-(q^2)}{f_+(q^2)}
\end{eqnarray}
The decay rate $\Gamma$ is well determined from experiment.
From Ref.~\cite{ref:ckm05:1},
\begin{equation}
| V_{us} f_+(0) | = 0.2173
\end{equation}
Hence, if we know $f_+(0)$ with extremely high accuracy, we can
determine $V_{us}$ with as much precision; this can impose a stringent
constraint on the unitarity of the CKM matrix.
The theoretical challenge on the lattice is to determine $f_+(0)$ with
such a high precision that the uncertainty is under control within
$\ll 1\%$.

Vector current conservation implies that $f_+(0) = 1$ in the exact
SU(3) limit of $m_u = m_d = m_s$.
The Ademello-Gatto theorem implies that the deviation of $f_+(0)$ from
unity is quadratic in the quark mass difference ($m_s - m_u$).
If we expand $f_+(0)$ in chiral perturbation theory
\cite{ref:bijnens:1},
\begin{equation}
f_+(0) = 1 + f_2 + f_4 + {\cal O}(p^6)
\end{equation}
where $f_i$ represents terms of order ${\cal O}(p^i)$.
Here, $f_2$ does not depend on any new low energy constants which can
be determined in terms of pseudoscalar meson masses and decay
constants.
However, $f_4$ does depends on new low energy constants.
The lattice technique to calculate $f_+(0)$ on the lattice
is the double ratio method \cite{ref:damir:1}.
\begin{equation}
\frac{\langle \pi | \bar{s} \gamma_4 u | K \rangle
\langle K | \bar{u} \gamma_4 s | \pi \rangle}
{\langle \pi | \bar{u} \gamma_4 u | \pi \rangle
\langle K | \bar{s} \gamma_4 s | K \rangle} =
\frac{(m_K + m_\pi)^2}{4 m_K m_\pi} | f_0(q^2_0) | ^2
\label{eq:f_0:dr}
\end{equation}
where the pion and kaon states carry zero momenta and $q_0^2 = (m_K -
m_\pi)^2$.
Here, note that $f_+(0) = f_0(0)$.
\begin{figure}[t!]
\begin{center}
\includegraphics[width=0.7\textwidth]{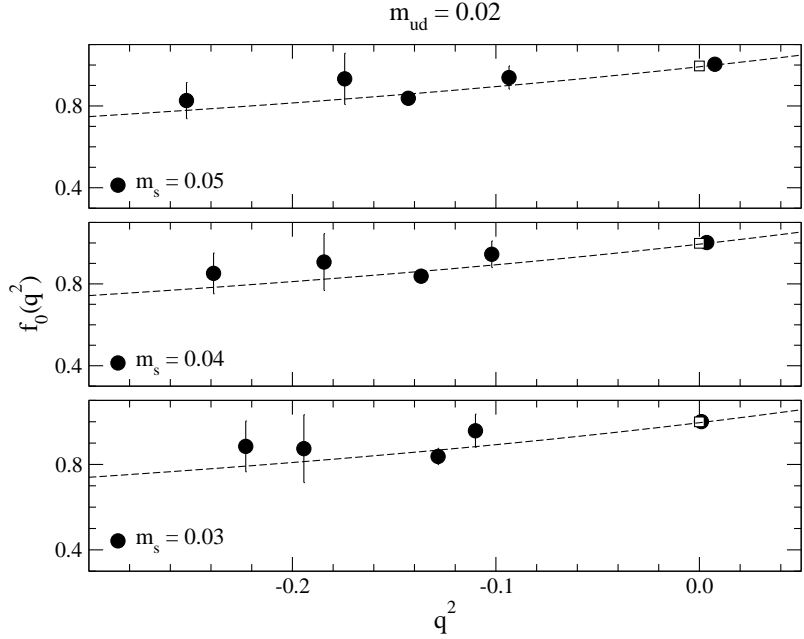}
\end{center}
\caption{$f_0(q^2)$ in two flavor QCD calculated by the RBC collaboration.}
\label{fig:f_0:rbc}
\end{figure}

Eq.~\ref{eq:f_0:dr} determines $f_0(q_0^2)$, but we need to know
$f_0(0)=f_+(0)$.
Hence, it is necessary to either extrapolate or interpolate in $q^2$ to
the (Minkowski) $q^2=0$ point at fixed pseudoscalar meson masses.
In other words, we need to insert small momenta in the kaon and pion
states to change $q^2 = (p-p')^2$ (defined in Minkowski space) from
$q^2_0 = (m_K - m_\pi)^2$ to negative $q^2$ such that we can
interpolate $f_0(q^2)$ to the $q^2=0$ point.
For this purpose, we can calculate the following ratio on the lattice:
\begin{eqnarray}
F(\vec{p},\vec{p}') &=& 
\frac{m_K + m_\pi}{E_K(\vec{p}) + E_\pi(\vec{p}')} \cdot
\frac{\langle \pi(\vec{p}') | \bar{s} \gamma_4 u | K(\vec{p}) \rangle}
{\langle \pi(\vec{0}) | \bar{s} \gamma_4 u | K(\vec{0}) \rangle}
\nonumber \\
&=& \frac{f_+(q^2)}{f_0(q^2_0)} \bigg(
1 + \frac{E_K(\vec{p}) - E_\pi(\vec{p}')}{E_K(\vec{p}) + E_\pi(\vec{p}')} 
\xi(q^2) \bigg)
\label{eq:F}
\end{eqnarray}
where $\vec{p}$ and $\vec{p}'$ are the kaon and pion momenta
respectively, $E_K(\vec{p})$ is the energy of kaon with a momentum
$\vec{p}$, and $E_\pi(\vec{p}')$ is the energy of pion with a momentum
$\vec{p}'$.
Here, note that $F(\vec{0},\vec{0}) = 1$.
In order to extract $f_0(q^2)$ from Eq.~\ref{eq:F}, we first need to
know $\xi(q^2)$.
It can be directly determined by measuring the following ratio:
\begin{eqnarray}
R(\vec{p},\vec{p}') &=&
\frac{\langle \pi(\vec{p}') | \bar{s} \gamma_i u | K(\vec{p}) \rangle
\langle K(\vec{p}') | \bar{s} \gamma_4 s | K(\vec{p}) \rangle}
{\langle \pi(\vec{p}') | \bar{s} \gamma_4 u | K(\vec{p}) \rangle
\langle K(\vec{p}') | \bar{s} \gamma_i s | K(\vec{p}) \rangle}
\nonumber \\
&=& 
\frac{[(p + p')_i + (p-p')_i \xi(q^2)]\cdot[E_K(\vec{p}) + E_K(\vec{p}')]}
{ [ (E_K(\vec{p}) + E_\pi(\vec{p}')) + 
(E_K(\vec{p}) - E_\pi(\vec{p}')) \xi(q^2)] 
\cdot [(p+p')_i]}
\label{eq:R}
\end{eqnarray}
After determining $\xi(q^2)$ from this, it is then possible to
determine $f_0(q^2)$ by combining Eq.~\ref{eq:f_0:dr}, \ref{eq:F} and
\ref{eq:R}.

Recently, a main goal on the lattice has been to calculate $f_0(0)$ in
unquenched QCD.
Three groups have reported results on this.
First, the RBC collaboration have calculated $f_0(q^2)$ in two flavor
QCD using domain wall fermions at $a = 0.12$ fm on the $16^3\times
32$ lattice \cite{ref:rbc:f_0}.
In Fig.~\ref{fig:f_0:rbc}, we show their results for $f_0(q^2)$ as a
function of (Minkowski) $q^2$.
Second, the JLQCD collaboration has calculated $f_0(q^2)$ in two
flavor QCD using non-perturbatively ${\cal O}(a)$ improved Wilson
fermions \cite{ref:jlqcd:f_0}.
In this calculation, the Wilson fermion action is ${\cal O}(a)$ improved
but the vector current operator is not fully ${\cal O}(a)$ improved.
Third, the HPQCD/FNAL collaboration has calculated $f_0(q^2)$ in $2+1$
flavor QCD \cite{ref:okamoto:f_0}.
They use unimproved staggered fermions for the $u$ and $d$ quarks and
an improved Wilson fermion for the $s$ quark.
They calculate only $f_0(q_0^2)$ using the double ratio in
Eq.~\ref{eq:f_0:dr} and extrapolate to $q^2=0$ using the pole model by
setting the free coefficient to the experimental value.
These results are summarized in Table \ref{tab:f_+:all}.
\begin{table}[t!]
\begin{center}
\begin{tabular}{ c | c }
\hline
collaboration & $f_+(0)$ \\
\hline
RBC        & $0.968(9)(6)$ ($N_F=2$)\\
HPQCD/FNAL & $0.962(6)(9)$ ($N_F=2+1$) \\
JLQCD      & $0.952(6)(-)$ ($N_F=2$) \\
Leutwyler  & $0.961(8)$ \\
\hline
\end{tabular}
\end{center}
\caption{Results for $f_+(0)$.}
\label{tab:f_+:all}
\end{table}

\section{Kaon distribution amplitude}
\label{sec:kda}
Exclusive reactions with specific hadrons in the final and initial
states have received a lot of attention.
The reason is that they are dominated by rare configurations of the
hadron's constituents.
One of these rare configurations is based on the hard mechanism in
which only valence quark configurations contribute and all quarks have
small transverse separation.
The structure of the hard mechanism is relatively simple to understand.
These hard contributions can be calculated in terms of hadron
distribution amplitudes (DA's) which describe the momentum fraction
distribution of partons at zero transverse separation with a fixed
number of constituents.
In particular, the leading-twist DA's of the pion and the nucleon have
been the focus of much attention.
The DA's of the kaon are important for understanding the exclusive $B$
decays such as $B\rightarrow K K^{*}$ in the framework of QCD
factorization, perturbative QCD and light cone sum rules.

The theoretical description of DA's for the kaon is based on the
relation:
\begin{equation}
\langle 0 | \bar{u}(z) \gamma_\nu \gamma_5 P(z,-z) s(-z)
| K^+(p) \rangle_{z^2 = 0} =
i f_K p_\nu \int^1_{-1} d\xi \ \exp(i\xi p \cdot z) \ \phi_K(\xi,\mu^2)
\label{eq:kda:1}
\end{equation}
where $z_\nu$ is a light-like vector, $z^2 = 0$, $\mu$ is a
renormalization scale, and $P(z,-z)$ is a path-ordered Wilson line
connecting the quark and the anti-quark fields in a gauge invariant
way:
\begin{equation}
P(z,-z) = {\cal P} \exp\bigg(-ig \int^{z}_{-z} dy^{\mu} A_\mu(y)\bigg)
\end{equation}
Here, $\phi_K$ is the leading twist DA of the kaon.
The physical interpretation of $\xi$ is that the $s$ quark carries an
$x=(1+\xi)/2$ fraction of the kaon momentum and the $\bar{u}$
anti-quark carries the rest (a $1-x = (1-\xi)/2$ fraction).
The $\phi_K$ parametrizes the overlap of the kaon state with
longitudinal momentum $p$ with the lowest Fock state consisting of a
quark and an anti-quark carrying the momentum fraction $xp$ and
$(1-x)p$, respectively.
In Eq.~\ref{eq:kda:1}, the convention for the normalization of
$\phi_K$ is as follows:
\begin{equation}
\int^1_{-1} d\xi \ \phi_K(\xi,\mu^2) = 1
\end{equation}

The conformal expansion provides a convenient tool to separate
the transverse modes and the longitudinal modes.
The transverse mode is formulated as a scale dependence of relevant
operators and the longitudinal mode is described in terms of 
Gegenbauer polynomials $C^{3/2}_n(\xi)$:
\begin{equation}
\phi_K(\xi,\mu^2) = \frac{3}{4}(1-\xi^2)
\bigg( 1 + \sum_{n=1}^{\infty} a_n(\mu^2) C^{3/2}_n(\xi) \bigg)
\end{equation}
We refer to the $a_n$'s as ``Gegenbauer moments''.

Applying the light-cone operator product expansion (OPE)
to the left-hand side of Eq.~\ref{eq:kda:1},
we can drive the following relationship for the $n$-th moment
$\langle \xi^n \rangle$ of the kaon DA:
\begin{eqnarray}
& & \langle \xi^n \rangle (\mu^2) =
\int^{1}_{-1} d\xi \ \xi^n \ \phi_K(\xi,\mu^2)
\\ & &
{\cal O}^{M}_{\mu_0,\mu_1,\cdots,\mu_n} =
i^n \bar{u}(0) \gamma_{\mu_0} \gamma_5
\stackrel{\leftrightarrow}{D}_{\mu_1} \cdots
\stackrel{\leftrightarrow}{D}_{\mu_n} s(0)
\label{eq:op:mink}
\\ & &
\langle 0 | {\cal O}^{M}_{\{\mu_0,\mu_1,\ldots,\mu_n\}} (0) | K(p) \rangle
= i f_K \ p_{\{\mu_0} \ldots p_{\mu_n\}} \ \langle \xi^n \rangle
\end{eqnarray}
where the operator is defined in Minkowski space,
$\stackrel{\leftrightarrow}{D} = \stackrel{\rightarrow}{D} -
\stackrel{\leftarrow}{D}$, and $\{\cdots\}$ denotes the traceless
symmetrization of all the indices.
The moments $\langle \xi^n \rangle$ are related to the Gegenbauer
moments $a_n$ as follows:
\begin{eqnarray}
a_1 &=& \frac{5}{3} \langle \xi \rangle \\
a_2 &=& \frac{7}{12} ( 5 \langle \xi^2 \rangle - 1 )
\end{eqnarray}

The theoretical mission on the lattice is to calculate the DA moments
$\langle \xi^n \rangle$ in Euclidean space.
The first step is to transcribe an operator in Euclidean space as
follows:
\begin{eqnarray}
{\cal O}^{E}_{\mu_0,\mu_1,\cdots,\mu_n} =
\bar{u}(0) \gamma_{\mu_0} \gamma_5
\stackrel{\leftrightarrow}{D}_{\mu_1} \cdots
\stackrel{\leftrightarrow}{D}_{\mu_n} s(0)
\end{eqnarray}
Since the symmetry group on the lattice is the finite group $H(4)$,
which is a subgroup of $O(4)$, we need to classify ${\cal
O}_{\mu_0,\mu_1,\cdots,\mu_n}$ in terms of irreducible representations
of $H(4)$.
The latter, in general, contains a number of irreducible
representations of $O(4)$.
For the first moment of the kaon, there are two types of operators:
\begin{eqnarray}
{\cal O}^a_{41} &=& \frac{1}{2} ( {\cal O}^E_{41} + {\cal O}^E_{14}) 
\\
{\cal O}^b_{44} &=& {\cal O}^E_{44} 
- \frac{1}{3} ({\cal O}^E_{11} + {\cal O}^E_{22} + {\cal O}^E_{33})
\end{eqnarray}
The ${\cal O}^a_{41}$ operator requires a nonzero momentum component
in the 1-direction.
The ${\cal O}^b_{44}$ operator can be evaluated at zero momentum.
\begin{figure}[t!]
\begin{center}
  \psfrag{msminusmq}[t][c][1][0]{$am_s-am_{ud}$}
  \psfrag{fstmom}[c][t][1][0]{$\langle \xi \rangle^{\rm bare}$}
  \psfrag{Legendmass1}[c][c][1][0]{$am_{ud}=0.01$}
  \psfrag{Legendmass2}[c][c][1][0]{$am_{ud}=0.02$}
  \psfrag{Legendmass3}[c][c][1][0]{$am_{ud}=0.03$}
  \includegraphics[height=0.8\textwidth,angle=270]{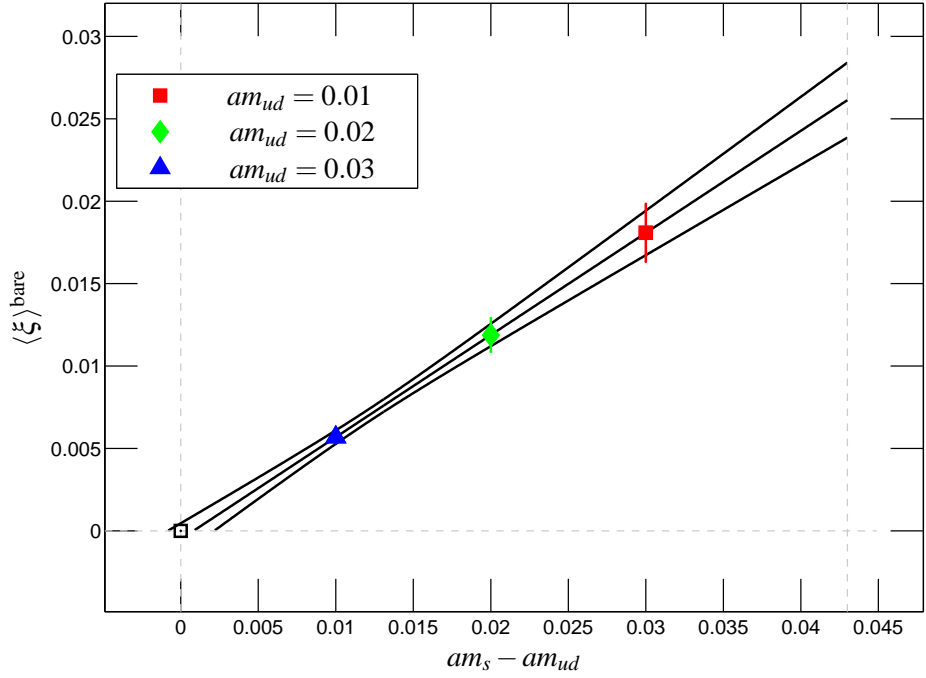}
\end{center}
\caption{$\langle \xi \rangle_K$ in $2+1$ flavor QCD calculated 
by the UKQCD collaboration using domain wall fermions.}
\label{fig:xi:ukqcd}
\end{figure}

Recently, the UKQCD collaboration has calculated the first moment of
the kaon DA in $2+1$ flavor QCD using domain wall fermions
\cite{ref:ukqcd:kda}.
They use the operator of ${\cal O}^a_{41}$ type.
In Fig.~\ref{fig:xi:ukqcd}, we show $\langle \xi \rangle_K$ as a function
of $m_s - m_{ud}$.
Chiral perturbation theory predicts that there is no correction with
chiral logarithms \cite{ref:chen:kda}.
The data is certainly consistent with this prediction. 
\begin{figure}[t!]
\begin{center}
  \includegraphics[width=0.8\textwidth]{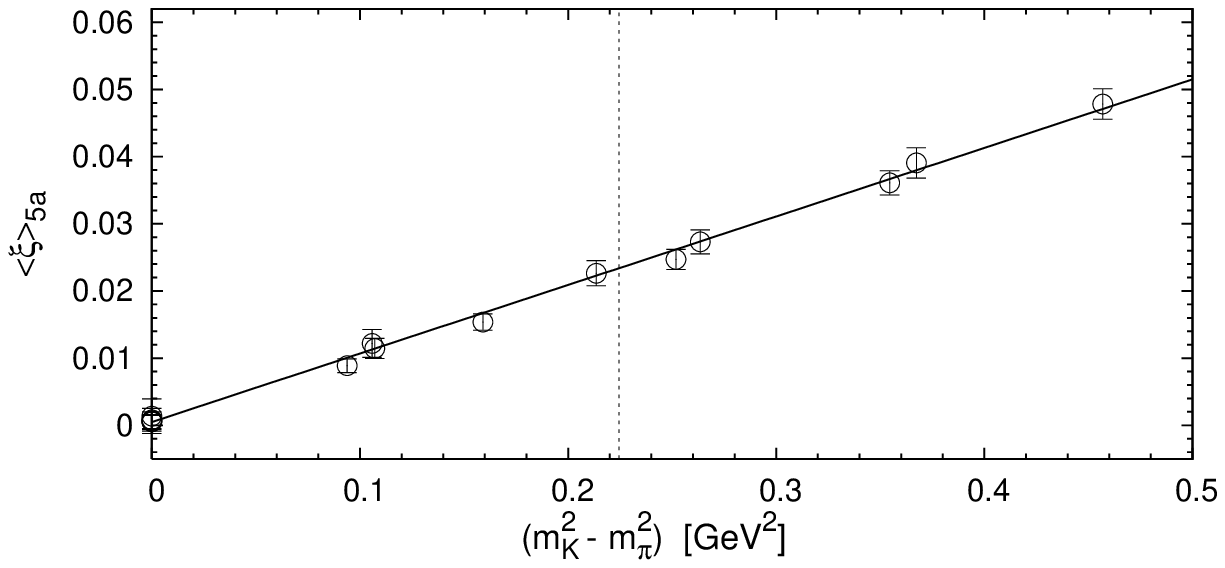}
  \includegraphics[width=0.8\textwidth]{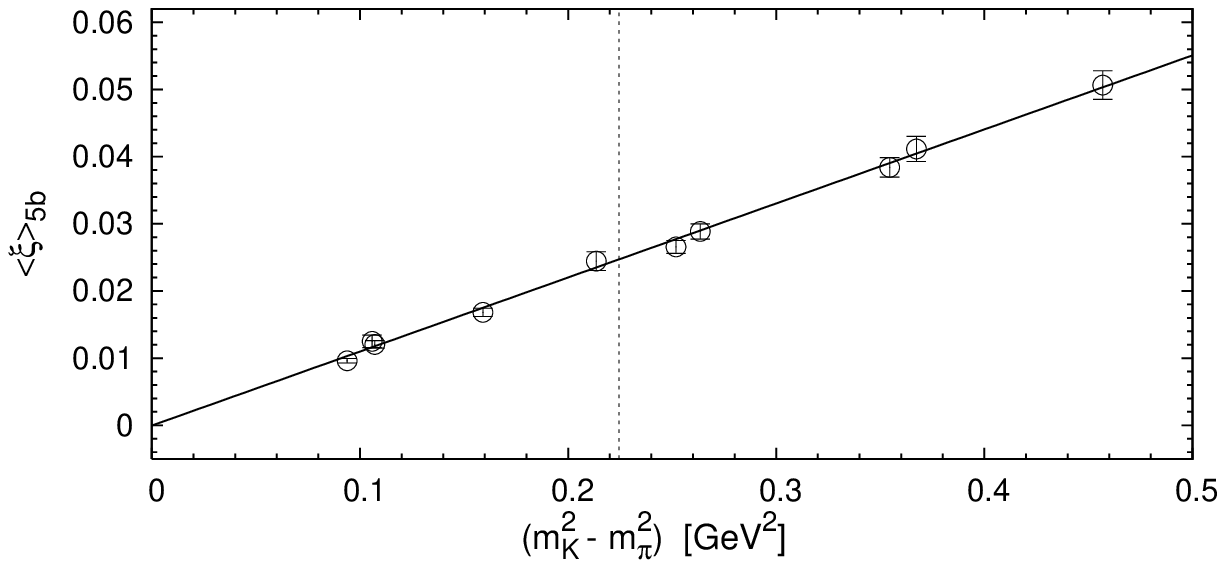}
\end{center}
\caption{$\langle \xi \rangle_K$ in two flavor QCD calculated by the
QCDSF/UKQCD collaboration using ${\cal O}(a)$ improved Wilson
fermions. Top: using ${\cal O}^a_{41}$. Bottom: using ${\cal
O}^b_{44}$.}
\label{fig:xi:qcdsf}
\end{figure}

The QCDSF/UKQCD collaborations have also calculated the first moment of
kaon DA in two flavor QCD using ${\cal O}(a)$ improved Wilson fermions
\cite{ref:qcdsf:kda}.
In Fig.~\ref{fig:xi:qcdsf}, we show $\langle \xi \rangle_K$ as a function
of $m_K^2 - m_{\pi}^2$.
In this calculation, they use operators of both ${\cal O}^a_{41}$ type
and ${\cal O}^b_{44}$ type.

\section{Summary and conclusion}
\label{sec:sum}
In this paper, we have reviewed recent progress in calculating kaon
spectrum, $B_K$, $\epsilon'/\epsilon$, $K\rightarrow \pi\pi$
amplitudes, kaon semileptonic form factors, and moments of the kaon
distribution amplitude on the lattice.
Especially, there has been much progress in calculating physical
observables in unquenched QCD using various lattice fermion
formulations.
For example, in the case of $B_K$, the quenched results of various
groups now agree with each other within statistical and systematical
uncertainty.

Through elaborate numerical studies on the pseudoscalar meson multiplet
spectrum and $B_K$, we learned that HYP staggered fermions are
remarkably better than AsqTad staggered fermions in reducing both
taste symmetry breaking effects and scaling violations.
However, HYP fat links are quite expensive when we calculate the
fermionic force term in the rational hybrid Monte Carlo (RHMC)
algorithm, because there are three SU(3) projection steps for each HYP
fat link.
For this reason, in Ref.~\cite{ref:wlee:2}, we proposed the
$\overline{\rm Fat7}$ scheme to improve staggered fermions for the
unquenched QCD simulation.
The $\overline{\rm Fat7}$ scheme shares the same advantage as the HYP
scheme in that the one-loop corrections for general composite
operators are identical in both schemes \cite{ref:wlee:2}.
In addition, the $\overline{\rm Fat7}$ scheme is computationally an
order of magnitude cheaper for the the unquenched QCD simulation using
the RHMC algorithm, mainly because we may have only one SU(3)
projection step for each $\overline{\rm Fat7}$ link and the memory
usage is only a quarter of that for the HYP scheme.
Taking into account all these facts, we recommend using the
$\overline{\rm Fat7}$ staggered fermions to generate unquenched gauge
configurations using the RHMC algorithm in the future.

There are a number of fundamental difficulties for finding a good
lattice version of the $Q_6$ and $Q_4$ operators in quenched QCD.
A solution to this particular problem has been proposed here.

In the case of direct calculation of $K\rightarrow \pi\pi$ amplitudes,
there have been a number of difficulties, in particular the Maiani
Testa no-go theorem.
A number of methods to get around the latter have been proposed and
tested extensively.
At present, the moving frame method looks quite promising and deserves
further investigation.
The twisted boundary condition method also looks promising but has not
been discussed extensively here due to lack of space.
It also deserves further attention in the future.

Most of the topics in kaon physics have their chiral behavior known
from (partially) quenched chiral perturbation theory.
However, the corresponding results in staggered chiral perturbation
theory are not available in many cases.
This also requires our attention and efforts in the future.

%
% EDIT
%

\section{Acknowledgments}
\label{sec:ack}
Proof-reading by David Adams is acknowledged with gratitude.
Helpful discussion with S.~Sharpe is also acknowledged with many
thanks.
W.~Lee acknowledges with gratitude that the research at Seoul National
University is supported by the KOSEF grant (R01-2003-000-10229-0), by
the KOSEF grant of international cooperative research program, by the
BK21 program, and by the DOE SciDAC-2 program.

\end{document}